\documentclass[showpacs,aps,twocolumn,floatfix,prb,longbibliography]{revtex4-2}
\usepackage{bm,amsmath,amssymb, mathrsfs}
\usepackage{bbm} %for unit matrix \mathbbm{1}
\usepackage{physics}
\usepackage{enumerate}
\usepackage{commath}
\usepackage{braket} %for Bra-Ket notation
\usepackage{graphicx,bm}
\usepackage{verbatim}
\usepackage[dvipsnames]{xcolor}
\usepackage{soul}
\usepackage[separate-uncertainty = true]{siunitx}
\usepackage[colorlinks]{hyperref}
\AtBeginDocument{%
    \hypersetup{    
        linkcolor=blue,
        filecolor=magenta,      
        urlcolor=blue,
        citecolor=blue,
     }
}

\newcommand{\eu}{\mathrm{e}}
\newcommand{\imi}{\mathrm{i}}

\usepackage[T1]{fontenc}
\usepackage[utf8]{inputenc}

\begin{document}

\title{Bulk spectra and the non-Hermitian skin effect in systems with long-range couplings}

\author{Jonathan Sturm}
\email{jonathan.sturm@uni-wuerzburg.de}

\author{Pia Kreß}

\author{Adriana P\'alffy}
\email{adriana.palffy-buss@uni-wuerzburg.de}

\affiliation{Julius-Maximilians-Universit{\"a}t W{\"u}rzburg, Institute for Theoretical Physics and Astrophysics and W{\"u}rzburg-Dresden Cluster of Excellence ctd.qmat, Am Hubland, 97074 W{\"u}rzburg, Germany}

\date{\today}% It is always \today, today,

\begin{abstract}
To achieve translational symmetry for the computation of the band structure of a lattice model, one can either consider an infinite lattice or impose periodic boundary conditions (PBC). 
While in systems with short-range couplings these two approaches are equivalent, we show that for long-range couplings one can obtain considerably different results.
We compare the two methods on the basis of one-dimensional quantum emitter chains both in free space and when coupled to a waveguide. 
The latter system allows for asymmetric couplings enabling the non-Hermitian skin effect, which we analyze using the two different band-structure calculation methods. 
We find that only PBC lead to physically and mathematically robust results, while the infinite-chain approach entails convergence issues and is unable to satisfyingly explain the emergence of the non-Hermitian skin effect in the waveguide system. 
Using PBC reveals unusual findings like a non-star shaped generalized Brillouin zone and strongly localized eigenstates with zero winding number.
\end{abstract}

\maketitle

\section{Introduction}
Band structures originally have been introduced in solid-state physics to study the bulk energy spectrum of electrons in a crystal while neglecting the system boundaries \cite{kittel2018}.
These bulk spectra and their corresponding Bloch states explain a vast number of phenomena, including the distinction between metals and insulators \cite{kittel2018} or the presence of protected boundary modes in topological insulators by means of the bulk-boundary correspondence \cite{chiu2016, Kunst2018, Wang2024review}. 
Over the decades, band structures have been adapted to many other platforms where one is interested in the physics of excitations in a periodic lattice, such as photonic crystals \cite{soukoulis_book}, electrical circuits \cite{Lee2018}, or quantum emitter lattices \cite{Garcia2017, Reitz2022, Sturm2025}.

One way to define and calculate the electronic band structure is the tight-binding model which considers the electrons to be tightly bound to their original atom and able to tunnel to their nearest neighbor by means of a finite overlap of the individual wavefunctions \cite{kittel2018}.
While in electronic systems this overlap decays fast with the distance between the atoms such that the coupling beyond nearest neighbors is negligible (``nearest-neighbor coupling''), in other platforms couplings often decay slower (``long-range couplings''). 
Examples include Rydberg atoms  with dipole-dipole couplings decaying with $1/r^3$ ($r$ being the distance between two atoms) \cite{Leseleuc2019, moegerle2026}, quantum emitters in free space ($1/r$ - $1/r^2$) \cite{Garcia2017, Garcia2017_A, Reitz2022}, helical Shiba chains ($1/r$) \cite{Pientka2013, Ren2026}, or even non-decaying couplings in waveguide-coupled quantum emitters \cite{Garcia2017, sheremet2023} or magnetic arrays \cite{Yu2022, Zeng2023}.
A recent review on systems with long-range couplings can be found in Ref.~\cite{defenu2023}.

While it is well understood how to define and calculate the band structure of tight-binding Hamiltonians with nearest-neighbor couplings, it turns out to be less obvious for systems with power-law decaying couplings. Two distinct methods have been used in the literature: one based on an infinite chain (in the following referred to as Method~I) and another one exploiting periodic boundary conditions (Method~II).
Although these two approaches yield equivalent results for short-range couplings, we show in this work that for long-range couplings they can lead to entirely different results.

This study presents clear definitions of the two Methods and compares them in a detailed analysis to determine which one yields a physically and mathematically meaningful band structure. 
We perform this analysis exemplarily using quantum emitter chains since not only they allow us to physically interpret our results (in contrast to abstract toy models); they also provide, depending on the exact implementation, various forms of long-range couplings, including non-reciprocal couplings which enable the non-Hermitian skin effect. 
Although Method~I has widely been used in the literature \cite{Albrecht2019, fedorovich2022, Calajo2022, sheremet2023, poddubny2024, Yu2022, Zeng2023, Garcia2017}, we find that for slowly or non-decaying couplings it can lead to severe physical and mathematical inconsistencies and it is incompatible with the established theories on the non-Hermitian skin effect. 
In contrast to this, Method~II is mathematically stable and physically consistent, making it in many situations the superior strategy. 
In our analysis of the non-Hermitian skin effect, Method~II also reveals peculiarities like a non-star shaped generalized Brillouin zone and eigenstates with zero winding that are yet strongly localized.

\section{Band structures under long-range couplings}\label{Sec II}
\begin{figure}
    \centering
    \includegraphics[width=1\linewidth]{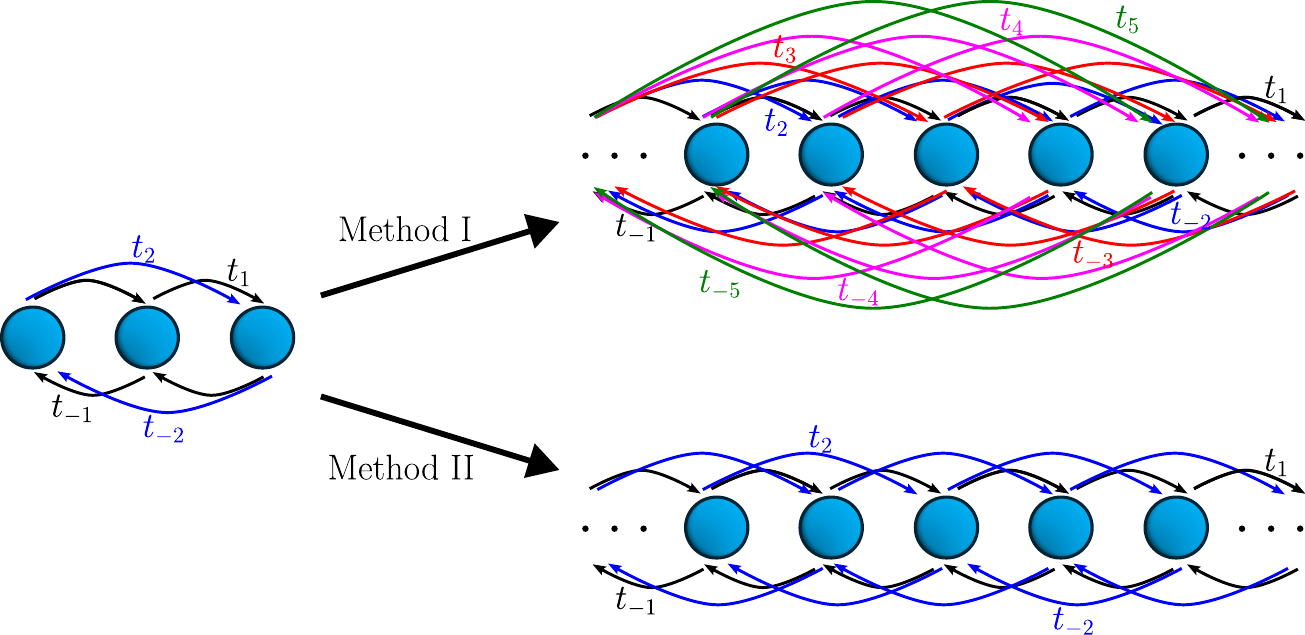}
    \caption{Illustration of the two Methods: A finite chain of $N=3$ sites with couplings $t_n$ ($-2\leq n \leq 2$) is extended to an infinite chain. 
    The infinite chain either incorporates all possible long-range couplings $t_n$, $n\in\mathbb{Z}$ (Method~I), or does only consider the coupling strengths present in the finite chain (Method~II).
    (For Method~I we only show a selection of couplings up to $|n|=5$ for illustrative purposes.)}
    \label{figure 1}
\end{figure}
Consider the generic tight-binding Hamiltonian
\begin{equation}
    H = \sum_{i,j=1}^N t_{ij} c_i^\dagger c_j,
\end{equation}
describing a one-dimensional chain of $N$ sites and a particle hopping from site $i$ to site $j$ with amplitude $t_{ij}\in\mathbb{C}$.
For our purposes, it is secondary whether the annihilation (creation) operators $c_i^{(\dagger)}$ are fermionic, bosonic, or else. 
Typically, in physical systems the couplings $t_{ij}$ are determined by the positions $\Vec{r}_i, \Vec{r}_j$ of the atoms $i,j$.
In case of translational symmetry, $t_{ij} = t_{i-j}\equiv t_n$ with $n=i-j$, we can also write the Hamiltonian as
\begin{equation}\label{eq hamiltonian initial}
    H = \sum_{i=1}^N \sum_{n=-i+1}^{N-i} t_n c_{i+n}^\dagger c_i.
\end{equation}
If there exist $p,q\in\mathbb{N}$ such that $t_n=0$ for all $n<-p$ and all $n>q$, we call the couplings $t_n$ ``short ranged'', otherwise ``long ranged''. 
Most standard models have $p=q=1$ (``nearest-neighbor couplings'').

In order to obtain the bulk spectrum of the chain, one extends the chain to infinite length, which removes the boundaries and leaves only bulk states, and diagonalizes the Hamiltonian with Bloch waves. 
In case of short-range couplings, this procedure is straight-forward and leads to the band structure \cite{Wang2024review, toeplitz_book}
\begin{equation}\label{eq omega short range}
    \omega(z) = \sum_{n=-p}^q t_n z^n,
\end{equation}
with $z=\eu^{\imi k}$, $-\pi\leq k\leq \pi$.
Conversely, the couplings $t_n$ can be recovered from the band structure as the Fourier coefficients \cite{toeplitz_book}
\begin{equation}\label{eq inverse couplings}
    t_n = \frac{1}{2\pi} \int_{-\pi}^\pi \omega(\eu^{\imi k}) \eu^{-\imi kn} \dd{k}.
\end{equation}
However, for long-range couplings there are at least two ways to extend the chain, which are illustrated in Fig.~\ref{figure 1}. 
One could either (\textbf{Method I}) consider an infinite chain with couplings $t_n$ generally non-zero for all $n\in\mathbb{Z}$, which leads to the band structure
\begin{equation}\label{eq omega 1}
    \omega_{\rm I} (z) = \sum_{n=-\infty}^\infty t_n z^n.
\end{equation}
Alternatively, one starts from the finite chain with length $N$ and extends this chain to infinite length while keeping the maximum coupling range $N-1$.
This approach (\textbf{Method II}) leads to the band structure
\begin{equation}\label{eq omega 2}
    \omega_{\rm II}(z) = \sum_{n=-(N-1)}^{N-1} t_n z^n.
\end{equation}
This is equivalent to imposing periodic boundary conditions on the finite chain, cf.~Appendix~\ref{sec PBC}.
Both Methods have been used in the literature, for instance, in Refs.~\cite{Albrecht2019, fedorovich2022, Calajo2022, sheremet2023, poddubny2024, Yu2022, Zeng2023, Garcia2017} (Method~I) and in Ref.~\cite{Reitz2022} (Method~II).
It is obvious that in general $\omega_{\rm I}(z) \neq \omega_{\rm II}(z)$ and that only in case of short-range couplings the two formulas lead to the same result if $p,q<N-1$.
For long-range couplings, the two formulas only coincide in the limit of an originally infinite chain, $\lim_{N\to\infty} \omega_{\rm II}(z) = \omega_{\rm I}(z)$.

In the following, we are going to compare these two Methods to determine which one yields a more accurate description of the bulk energies of a finite chain.

\section{Systems with symmetric couplings}\label{sec III}
In order to compare our two Methods, we are going to study a chain of quantum emitters as they allow us to study both symmetric ($t_n=t_{-n}$) and asymmetric ($t_n \neq t_{-n}$), or chiral couplings and various dependencies of $t_n$, from fast decaying couplings $t_n\sim 1/n^2$ in free space to non-decaying couplings in a waveguide system.
Asymmetric couplings will also allow us to study the non-Hermitian skin effect and its bulk-boundary correspondence based on the two types of band structures in Sec.~\ref{sec nhse}.

\subsection{Example 1: quantum emitters in free space}
Quantum emitters can be described as two-level systems with ground and excited states $\ket{\rm g}$ and $\ket{\rm e}$, respectively, which can be excited by absorption of a photon with resonant frequency $\omega_0$. 
Two quantum emitters in close vicinity can exchange excitation via the surrounding electromagnetic field. 
For a chain of $N$ identical quantum emitters in free space, the effective Hamiltonian reads \cite{Garcia2017, Reitz2022}
\begin{equation}\label{eq quantum emitter Hamiltonian}
    H = \sum_{i=1}^N \sum_{n=-i+1}^{N-i} t_n \sigma_{i+n}^\dagger \sigma_i,
\end{equation}
where $\sigma_i = \ketbra{\mathrm{g}_i}{\mathrm{e}_i}$ and \cite{Garcia2017, Garcia2017_A}
\begin{align}\label{eq free space couplings}
    t_n = -\Gamma_0 \frac{3}{4}\eu^{\imi\varphi|n|} \Bigg( & \qty(\frac{1}{\varphi|n|} + \frac{\imi}{(\varphi|n|)^2} - \frac{1}{(\varphi|n|)^3}) \nonumber \\
    & - \qty(\frac{1}{\varphi|n|} + \frac{3\imi}{(\varphi|n|)^2} - \frac{3}{(\varphi|n|)^3}) \beta\Bigg)
\end{align}
for $n\neq 0$ and $t_0 = \omega_0 -\imi\frac{\Gamma_0}{2}$.
Here, $\Gamma_0$ is the single-atom spontaneous decay rate, $\varphi = k_0 a$ with $k_0 = \frac{\omega_0}{c}$ being the wavenumber of the atomic transition ($c$ is the speed of light) and $a$ the lattice spacing, i.e., the distance between two neighboring emitters. 
The value of $\beta$ depends on the polarization of the chain (i.e., the orientation of the individual transition dipole moments with respect to the chain axis) and is given by $\beta=1$ ($\beta=0$) for parallel (perpendicular) polarization. 
Throughout this work, we are working in the single-excitation basis given by $\ket{i}=\sigma_i^\dagger\ket{\rm GS}$, where $i=1,...,N$ and $\ket{\rm GS}$ is the collective ground state.

In the following, we derive the band structures of this system using the two Methods.
The results for Method~I have been obtained in Ref.~\cite{Garcia2017}, while Method~II was used in Ref.~\cite{Reitz2022}.

\subsubsection{Results for Method I}
\textit{Parallel polarization.}
In this case, $\beta=1$ and we have for $n\neq 0$
\begin{equation}
    t_n = \Gamma_0 \frac{3}{2} \eu^{\imi\varphi|n|} \qty(\frac{\imi}{(\varphi|n|)^2} - \frac{1}{(\varphi|n|)^3}).
\end{equation}
Since the series $\sum_{n=1}^\infty \frac{1}{n^2}$ converges, the band structure in Eq.~\eqref{eq omega 1} is well-defined and reads \cite{Garcia2017}
\begin{align}
    \frac{\omega_{\rm I}(k)}{\Gamma_0} = \frac{3}{2\varphi^3}
        \Bigg( & \imi\varphi \mathrm{Li}_2(\eu^{\imi(k_0-k)a}) + \imi\varphi \mathrm{Li}_2(\eu^{\imi(k_0+k)a}) \nonumber \\
        & - 3\mathrm{Li}_3(\eu^{\imi(k_0-k)a}) - 3\mathrm{Li}_3(\eu^{\imi(k_0+k)a})\Bigg) \nonumber \\
        & + \frac{\omega_0}{\Gamma_0} - \frac{\imi}{2},
\end{align}
where $\mathrm{Li}_s(z) = \sum_{n=1}^\infty \frac{z^n}{n^s}$ is the polylogarithm.
Here and in the following text we often write $\omega_{\rm I/II}(k) \equiv \omega_{\rm I/II}(\eu^{\imi ka})$. \\

\textit{Perpendicular polarization.}
In this case, $\beta=0$, and therefore for $n\neq 0$
\begin{equation}\label{eq free space perendicular couplings}
    t_n = -\Gamma_0\frac{3}{4}\eu^{\imi\varphi|n|} \qty(\frac{1}{\varphi|n|} + \frac{\imi}{(\varphi|n|)^2} - \frac{1}{(\varphi|n|)^3}).
\end{equation}
For $k\neq\pm k_0$, the Laurent series in Eq.~\eqref{eq omega 1} converges, leading to \cite{Garcia2017}
\begin{align}\label{eq omega I freespace perpendicular}
    \frac{\omega_{\rm I}(k)}{\Gamma_0}
    = 
    -\frac{3}{4} \Bigg( & \frac{1}{\varphi} \mathrm{Li}_1(\eu^{\imi(k_0+k)a}) + \frac{1}{\varphi} \mathrm{Li}_1(\eu^{\imi(k_0-k)a}) \nonumber \\
    & + \frac{\imi}{\varphi^2} \mathrm{Li}_2(\eu^{\imi(k_0+k)a}) + \frac{\imi}{\varphi^2} \mathrm{Li}_2(\eu^{\imi(k_0-k)a}) \nonumber \\
    & - \frac{1}{\varphi^3} \mathrm{Li}_3(\eu^{\imi(k_0+k)a}) - \frac{1}{\varphi^3} \mathrm{Li}_3(\eu^{\imi(k_0-k)a}) \Bigg) \nonumber \\
    & +\frac{\omega_0}{\Gamma_0} - \frac{\imi}{2}.
\end{align}
However, for $k=\pm k_0$, the $(1/n)$ term in Eq.~\eqref{eq free space perendicular couplings} leads to a harmonic series, $\omega_{\rm I}(\pm k_0) \sim \sum_{n=1}^\infty \frac{1}{n} = \infty$, which is known for its divergence. 
These divergencies can also be seen in the first line of Eq.~\eqref{eq omega I freespace perpendicular} as $\mathrm{Li}_1(z) = -\ln(1-z)$.

\subsubsection{Comparison with Method II}
\begin{figure}
    \centering
    \includegraphics[width=1\linewidth]{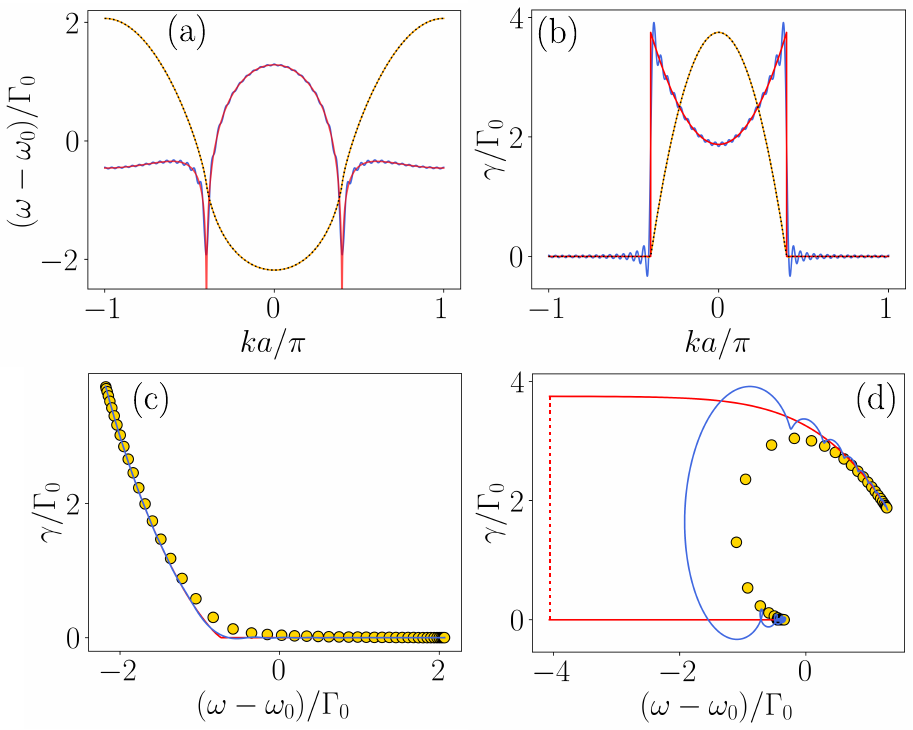}
    \caption{Band structures of a quantum emitter chain in free space. 
    (a) and (b) show the frequency shifts $\omega=\mathrm{Re}(\omega_{\rm I/II}(k))$ and collective decay rates $\gamma = -2\mathrm{Im}(\omega_{\rm I/II}(k))$. 
    Parallel polarization is shown in the black dashed (Method~I) and orange (Method~II) curves, perpendicular polarization in the red (Method~I) and blue (Method~II) curves.
    (c) and (d) show the curves $\omega_{\rm I/II}$ (red/blue) as well as the finite size eigenvalues (yellow dots) in the complex plane for parallel and perpendicular polarization, respectively. 
    The red dashed line in (d) indicates the discontinuity that actually takes place at $\omega/\Gamma_0 =-\infty$.
    Parameters in all panels are $N=50$ and $\varphi=2\pi/5$.}
    \label{figure 2}
\end{figure}
In Fig.~\ref{figure 2}, we compare the analytical results for $\omega_{\rm I}$ (Method~I) with the numerical evaluation of Eq.~\eqref{eq omega 2} (Method~II) for a chain of $N=50$ quantum emitters.
Figure~\ref{figure 2}(a) shows the collective dipole shifts $\omega(k) = \mathrm{Re}(\omega_{\rm I/II}(k))$ as a function of momentum $k$ for both types of polarization. 
While the curves for parallel polarization perfectly match (black dotted $\hat{=}$ Method~I, orange $\hat{=}$ Method~II), for perpendicular polarization there are visible discrepancies. 
The red curve (Method~I) clearly shows the two divergencies at $k=\pm k_0$ mentioned above, which originate from the non-convergence of the harmonic series, while the blue curve (Method~II) is smooth around these points due to the finiteness of the sum in Eq.~\eqref{eq omega 2}.
Moreover, the finite sum leads to small oscillations of the blue curve around the red curve, which get smaller for increasing $N$, see also Fig.~5(b) in Ref.~\cite{Reitz2022}.
Figure~\ref{figure 2}(b) shows the collective decay rates $\gamma(k) = -2\mathrm{Im}(\omega_{\rm I/II}(k))$. 
Again, we observe perfect agreement for parallel polarization, while for perpendicular polarization the curves differ by the small oscillations of the blue curve. 
We note that the non-physical oscillations around $\gamma=0$ for $|k|>k_0$, which lead to negative decay rates, are an artifact of Method~II known as Gibbs phenomenon \cite{Hewitt1979} since we diagonalize an infinite chain while only considering finite-range couplings.
(Equivalently, one could argue that these artifacts stem from additional couplings when imposing periodic boundary conditions, cf. Appendix~\ref{sec PBC}.) 
Despite this, the collective decay rate is still normalized as
\begin{equation}\label{eq normalization decay rate}
    \frac{1}{2\pi} \int_{-\pi/a}^{\pi/a} \gamma(k) \dd{k} = \Gamma_0.
\end{equation}
Moreover, since for a finite system the collective decay rates will always be positive, these small negative oscillations are benign artifacts.

In Figs.~\ref{figure 2}(c) and (d), the curves $\omega_{\rm I/II}(k)$ (red / blue) are shown in the complex plane. 
To compare the band structures with a finite chain, we also show the eigenvalues of the Hamiltonian in Eq.~\eqref{eq quantum emitter Hamiltonian} for $N=50$ (yellow dots). 
While the two curves and the discrete eigenvalues almost perfectly agree for parallel polarization [panel~(c)], for perpendicular polarization [panel~(d)] major differences can be observed.
The red curve (Method~I) has a discontinuity (dashed line) at which it jumps from $\gamma=4\Gamma_0$ to $\gamma=0$, corresponding to the divergencies in Fig.~\ref{figure 2}(a) at $k=\pm k_0$. 
The position of the discontinuity on the real axis depends on the numerical fineness of the $k$ grid and lies further left the finer the grid is. 
(Analytically, the jump happens at $\omega/\Gamma_0 = -\infty$.)
Considering the finite system eigenvalues (yellow dots), there is a stark contrast to the red band structure. 
Only at the start and end points the dots coincide with the curve, while the discrepancy becomes most obvious around the discontinuity. 
The eigenvalues follow a smooth arc with dense accumulations at the starting and ending points. 
This behavior is well reflected in the Method~II band structure, which follows a similar arc with slightly larger radius and its characteristic oscillations, which become faster at the starting and ending points of the curve analogously to the eigenvalue accumulations.

To summarize this first example, we can conclude that for parallel polarization, the two Methods yield identical results, while for perpendicular polarization there are some discrepancies: the small oscillations in Method~II vs. the smooth Method~I curve and the divergencies in Method~I that are absent in Method~II. 
From that, one may infer that the two Methods agree the better the faster the couplings decay (parallel polarization has a $1/n^2$ dependence, while perpendicular polarization leads to the $1/n$ decay and the diverging harmonic series).
Therefore, we can suspect that the situation will become even more interesting when looking at a model with non-decaying couplings.

\subsection{Example 2: waveguide model}
Quantum emitters are not only studied in free space but also within structured electromagnetic environments tailoring the dipole-dipole couplings. 
One prominent example of such a system is a chain of quantum emitters coupled by the guided mode of a waveguide, which is resonant to the atomic transition, see Ref.~\cite{sheremet2023} for a recent review. 
Not only enables this system non-decaying long-range couplings, but also directional couplings \cite{Lodahl2017}, which are a necessary ingredient for the non-Hermitian skin effect, which we will study in Sec.~\ref{sec nhse}. 
The Hamiltonian of a quantum emitter chain coupled to a waveguide has the same form as in Eq.~\eqref{eq quantum emitter Hamiltonian} with the couplings now being \cite{fedorovich2022, sheremet2023, poddubny2024}
\begin{equation}\label{eq couplings waveguide}
    t_n = -\imi \eu^{\imi \varphi |n|} \times 
    \begin{cases}
        \gamma_{\rm L} & (n<0) \\
        \gamma_{\rm R} & (n>0)
    \end{cases}
\end{equation}
for $n\neq 0$ and $t_0 = \omega_0 - \imi\frac{\gamma_0}{2}$.
Here, $\gamma_0$ is the rate on which a single atom decays into the guided mode (therefore the distinction to the previous $\Gamma_0$) and $\varphi = k_0 a$, where $a$ is the lattice constant and $k_0$ the guided mode wave number. 
The directional coupling strengths are parametrized by the directionality parameter $\xi = \gamma_{\rm R}/\gamma_{\rm L}$ as \cite{fedorovich2022, sheremet2023, poddubny2024}
\begin{equation}\label{eq waveguide gammas}
    \gamma_{\rm L} = \gamma_0 \frac{1}{1+\xi}, \ \ \ \gamma_{\rm R} = \gamma_0 \frac{\xi}{1+\xi} \ \ \ (0\leq \xi \leq 1).
\end{equation}
For $\xi=1$, the couplings become symmetric, $\gamma_{\rm L} = \gamma_{\rm R}$, while for $\xi=0$ the couplings become fully chiral toward the left.

This model has been studied using Method~I in Refs.~\cite{fedorovich2022, sheremet2023, poddubny2024, Calajo2022, Albrecht2019}, while Refs.~\cite{Yu2022, Zeng2023} discuss a similar Hamiltonian in magnetic arrays. 
For Method~II, there are no previous results that we are aware of.

\subsubsection{Results for Method I}
Inserting the couplings from Eq.~\eqref{eq couplings waveguide} into Eq.~\eqref{eq omega 1}, we obtain the series
\begin{equation}\label{eq omega 1 waveguide series}
    \omega_{\rm I}(k) = -\imi\gamma_{\rm L}\sum_{n=1}^\infty q_-^n -\imi\gamma_{\rm R} \sum_{n=1}^\infty q_+^n + \left(\omega_0 - \imi\frac{\gamma_0}{2}\right)
\end{equation}
with $q_\pm = \eu^{\imi(k_0\pm k)a}$, where one can identify the geometric series
\begin{equation}
    \sum_{n=1}^\infty q^n = \frac{q}{1-q}.
\end{equation}
However, the geometric series only converges for $|q|<1$ and diverges for $|q|\geq 1$. 
Since in our case $|q|=1$, the series diverges and the band structure $\omega_{\rm I}$ is ill-defined.
If one still uses the above formula, one obtains \cite{fedorovich2022, poddubny2024, sheremet2023}
\begin{equation}\label{eq omegaI waveguide}
    \omega_{\rm I}(k)
    = \omega_0 + \frac{\gamma_0}{2(1+\xi)}
    \qty(\cot(\varphi_-/2) + \xi\cot(\varphi_+/2))
\end{equation}
with $\varphi_\pm = (k_0\pm k)a$.
Similar to the free-space situation, there are two divergencies at $k=\pm k_0$.

Although the series in Eq.~\eqref{eq omega 1 waveguide series} does not immediately converge as it is written, there are ways to make the series converge.
To obtain classical convergence (e.g., pointwise convergence), one needs to add a finite decay to the couplings such that $|q_\pm|<1$.
This approach changes the system physically and will be discussed in Sec.~\ref{sec Skin effect with method I}.
Without additional decay, the series can only converge in a generalized sense (e.g., distributionally, leading to a Dirac comb).
Since such band structures are unphysical, we will give them no further consideration.

\subsubsection{Results for Method II}
With Method II, the band structure reads
\begin{equation}
    \omega_{\rm II}(k)
    = -\imi\gamma_{\rm L} \sum_{n=1}^{N-1} q_-^n - \imi\gamma_{\rm R} \sum_{n=1}^{N-1} q_+^n + \left(\omega_0 - \imi\frac{\gamma_0}{2}\right).
\end{equation}
Due to the finiteness of the sums, we will not encounter any convergence problems and can deliberately use the formula of the geometric sum 
\begin{equation}
    \sum_{n=1}^{N-1} q^n = \frac{q - q^N}{1 - q}
\end{equation}
to obtain
\begin{equation}\label{eq omegaII waveguide}
    \omega_{\rm II}(k)
    = \omega_{\rm I}(k) - \frac{\gamma_0}{2(1+\xi)} 
    \qty(\frac{\eu^{\imi\varphi_-(N-\frac{1}{2})}}{\sin(\varphi_-/2)} + \xi \frac{\eu^{\imi\varphi_+(N-\frac{1}{2})}}{\sin(\varphi_+/2)}).
\end{equation}
In contradiction to our previous statement that in general we have $\lim_{N\to\infty} \omega_{\rm II} = \omega_{\rm I}$, cf. Eq.~\eqref{eq omega 2} and below, one can see that this result does not converge to $\omega_{\rm I}$ for $N\to\infty$, but instead increasing $N$ leads to faster rotating terms, confirming the non-convergence of the geometric series in Eq.~\eqref{eq omega 1 waveguide series}.

\subsubsection{Comparison}
\begin{figure}
    \centering
    \includegraphics[width=1\linewidth]{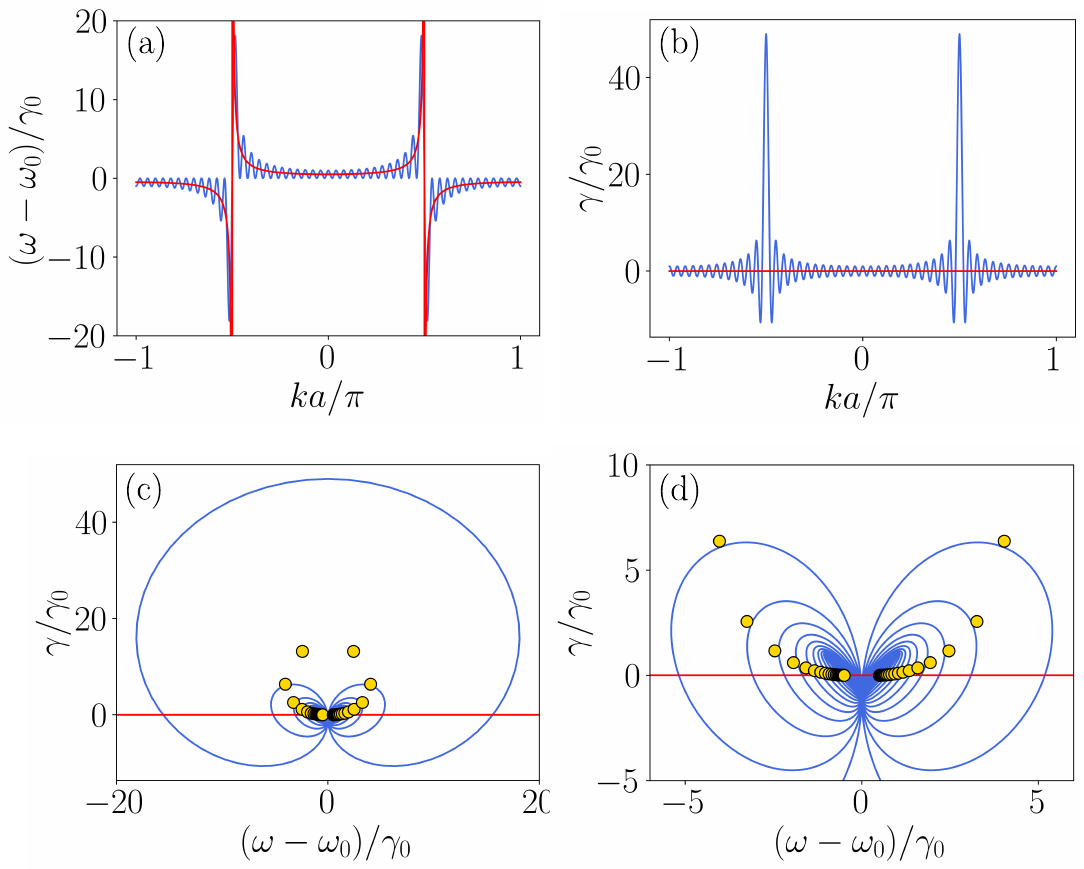}
    \caption{Band structures of a quantum emitter chain coupled to a waveguide. 
    (a) and (b) show the frequency shifts $\omega=\mathrm{Re}(\omega_{\rm I/II})$ and collective decay rates $\gamma = -2\mathrm{Im}(\omega_{\rm I/II})$, respectively. 
    The red (blue) curves show the results for Method~I~(II).
    (c) shows the band structures in the complex plane with finite chain eigenvalues (yellow dots). 
    (d) shows a zoom of (c).
    Parameters in all panels are $N=50$, $\varphi=\pi/2$, and $\xi=1$.}
    \label{figure 3}
\end{figure}
In Fig.~\ref{figure 3}, we compare the band structures $\omega_{\rm I}$ (red) and $\omega_{\rm II}$ (blue) for $\xi=1$, which is the case of zero chirality, and $\varphi=\pi/2$.
We will discuss chiral couplings, which lead to the non-Hermitian skin effect, in Sec.~\ref{sec nhse}.
Different values of $\varphi$ lead to a slightly asymmetric spectrum without affecting the qualitative results \cite{poddubny2024}, which is why we can restrict our analysis to the case $\varphi=\pi/2$.
Figure~\ref{figure 3}(a) shows the frequency shift $\omega(k) = \mathrm{Re}(\omega_{\rm I/II}(k))$.
The red curve contains the two divergencies at $k=\pm k_0$ as discussed above and is smooth otherwise. 
The blue curve does not diverge and shows the characteristic oscillations around the red curve, leading to smooth transitions at $k=\pm k_0$.
While the real parts match fairly well, the collective decay rates $\gamma(k) = -2\mathrm{Im}(\omega_{\rm I/II}(k))$ in Fig.~\ref{figure 3}(b) show severe disagreements.
Since the result for Method~I in Eq.~\eqref{eq omegaI waveguide} is purely real, collective decay rates vanish for all $k$, which necessarily violates the normalization condition in Eq.~\eqref{eq normalization decay rate}. 
In contrast to this, the collective decay rates obtained with Method~II in Eq.~\eqref{eq omegaII waveguide} are finite with negative amplitudes due to oscillations but conserved normalization, similar to the free-space case.

Figure~\ref{figure 3}(c) shows the band structures in the complex plane. 
While $\omega_{\rm I}$ (red curve) appears as a straight line covering the real axis due to the vanishing collective decay rate and the divergence of the real part, the oscillations of $\omega_{\rm II}$ lead to the formation of multiple loops, the origin of which are the $\eu^{\imi \varphi_\pm N}$ terms in Eq.~\eqref{eq omegaII waveguide}. 
Therefore, the number of loops increases with $N$. 
We need to mention that, despite its appearance, the blue curve in Fig~\ref{figure 3}(c) is \textit{not} a closed loop. 
Although the two ends of the ribbon seem to touch each other around $(\omega-\omega_0, \ \gamma) = (0, \ -\gamma_0)$, see the zoom in Fig.~\ref{figure 3}(d), the curve is double covered as $\omega_{\rm II}(k) = \omega_{\rm II}(-k)$, as expected for a system with symmetric couplings.

\section{Non-Hermitian skin effect}\label{sec nhse}
\begin{figure*}
    \centering
    \includegraphics[width=1\linewidth]{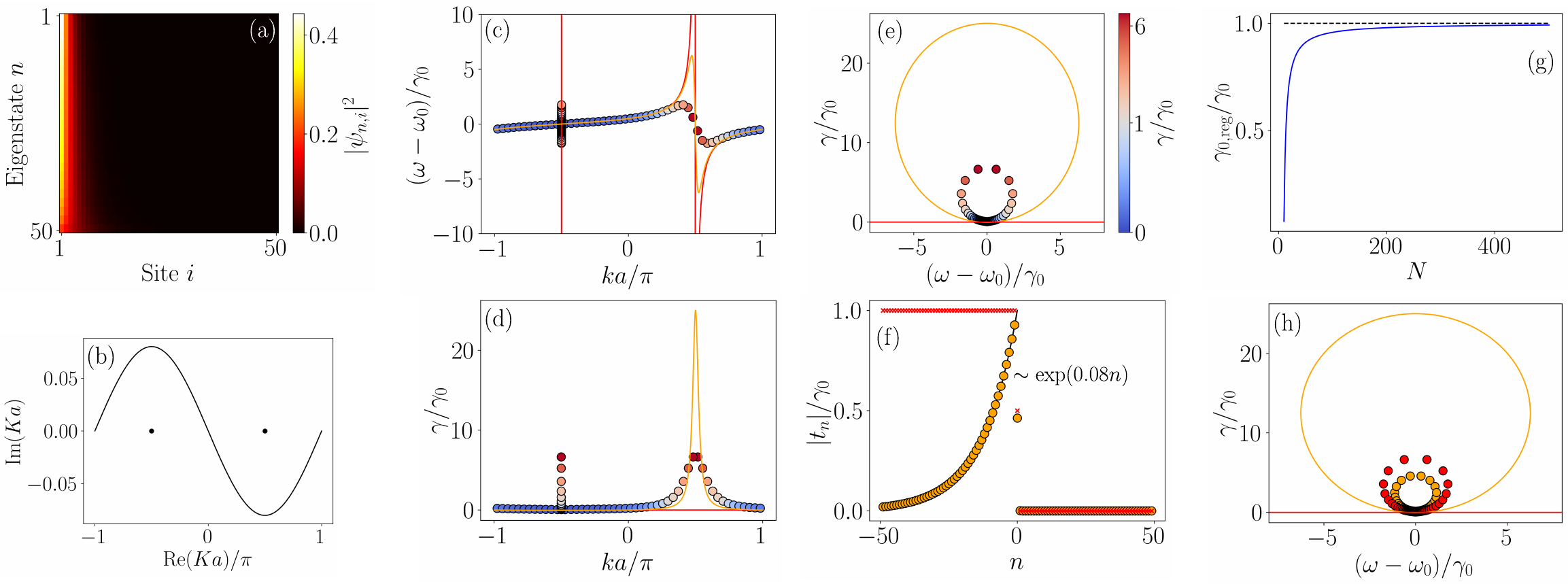}
    \caption{Non-Hermitian skin effect in a quantum emitter chain chirally coupled to a waveguide with directionality parameter $\xi=2\times 10^{-5}$.
    (a) Localization of eigenstates $\ket{\psi_n}$ of the finite size Hamiltonian in Eq.~\eqref{eq quantum emitter Hamiltonian} with $N=50$. 
    Eigenstates are sorted by the negative imaginary part of their corresponding eigenvalues (i.e., $n=1$ corresponds to highest decay rate $\gamma$, $n=N$ to the lowest).
    (b) Regularization path according to Eq.~\eqref{eq regularization}.
    (c) and (d) show the frequency shift $\omega=\mathrm{Re}(\omega_{\rm I})$ and collective decay rates $\gamma = -2\mathrm{Im}(\omega_{\rm I})$, respectively, as functions of $k$. 
    The red (orange) curve is the original (regularized) band structure.
    Colored dots are finite size eigenvalues [color bar in panel~(e)].
    (e) Band structures and finite size eigenvalues in the complex plane.
    (f) Couplings $|t_n|$ of the original model in Eq.~\eqref{eq couplings waveguide} (red crosses) and of the regularized band structure (orange dots). 
    The black curve indicates the fitted exponential decay of the couplings.
    For positive $n$, red crosses and orange dots almost perfectly overlap.
    (g) Single emitter spontaneous decay rate $\gamma_{\rm reg}$ after regularization as function of chain length $N$.
    (h) Original and regularized band structures (red and orange curves, respectively) as well as original and regularized eigenvalues (red and orange dots, respectively). 
    Parameters used for all graphs are $N=50$ [except for panel~[g]), $\xi=2\times10^{-5}$, and $\varphi=\pi/2$.}
    \label{figure 4}
\end{figure*}
The non-Hermitian skin effect describes the formation of localized modes at one of the boundaries of a finite-size chain due to asymmetric couplings, $t_n \neq t_{-n}$ \cite{Wang2024review}.
Such asymmetry can be achieved in the waveguide system from the previous section by choosing a finite chirality, $\xi<1$, as in this case $\gamma_{\rm L} > \gamma_{\rm R}$, cf. Eq.~\eqref{eq waveguide gammas}.
The calculated eigenstate localization for the case $\xi=2\times10^{-5}$ is shown in Fig.~\ref{figure 4}(a).
The emergence of skin modes is linked to the non-trivial point gap topology of the band structure $\omega$, which can be derived from the spectral winding number \cite{Wang2024review, Okuma2020, zhang2020winding}
\begin{equation}
    \nu(\varepsilon_0) = \frac{1}{2\pi\imi} \int_{-\pi}^{\pi} \dv{k} \ln(\omega(k) - \varepsilon_0) \dd{k},
\end{equation}
that counts the number of times the curve $\omega(k)$ encircles a point $\varepsilon_0\in\mathbb{C}$ counterclockwise.

In the following, we analyze the non-Hermitian skin effect in the waveguide system introduced in Sec.~\ref{sec III} based on the two investigated Methods to calculate the band structure. We discuss their ability to explain the eigenstate localization on the basis of non-trivial point gap topologies.

\subsection{Skin effect with Method I}\label{sec Skin effect with method I}
We begin by shortly summarizing and complementing the results of Ref.~\cite{poddubny2024}, 
which investigates the non-Hermitian skin effect  in the waveguide model using  Method~I to calculate the band structure.
The main challenge with the band structure obtained in Eq.~\eqref{eq omegaI waveguide} is that it is purely real, implying $\nu(\varepsilon)=0$ for any $\frac{\varepsilon}{\gamma_0}\in\mathbb{C}$.
Therefore, at first, the Method~I band structure has a trivial point gap topology, which could not explain the eigenstate localization in Fig.~\ref{figure 4}(a).

In order to both get rid of the divergencies at $k=\pm k_0$ and make $\omega_{\rm I}$ form an actual loop in the complex plane, Ref.~\cite{poddubny2024} suggests to regularize the band structure, which can be done by replacing \cite{poddubny2024} 
\begin{equation}\label{eq regularization}
    k \mapsto K(k) = k - \frac{4\imi}{N a} \sin(ka),
\end{equation}
where $N$ is the number of emitters in the finite chain. 
The regularized $K$ is shown in Fig.~\ref{figure 4}(b).

Figure~\ref{figure 4}(c) shows the real part of the original band structure $\omega_{\rm I}(k)$ (red curve) for $\xi=2\times10^{-5}$ alongside the regularized curve (orange) and the finite system eigenvalues shown as dots. 
Let us first focus on the original band structure (red curve).
Also here we can identify the two divergencies, however the divergency at $k=+k_0$ has become visibly broader compared to the non-chiral case [cf. Fig.~\ref{figure 3}(a)], while the one at $k=-k_0$ has become extremely narrow. 
We need to point out that the narrow divergency at $k=-k_0$ has not been shown in previous works \cite{sheremet2023, fedorovich2022}, possibly because of the extreme sharpness of the peak requiring a very fine grid for the $k$ values when evaluating $\omega_{\rm I}(k)$. 
Therefore, one can easily oversee this peak in the numerical analysis, although the analytical result in Eq.~\eqref{eq omegaI waveguide} unambiguously postulates its existence. 
We can also see that the finite size eigenvalues actually follow the red curve at this point, leading to an (almost perfectly) vertical pile of eigenvalues at $k=-k_0$.
(We discuss the mapping from eigenvalues to $k$ values in Appendix~\ref{appendix eigenvalue mapping}.)
In contrast to the red curve, the regularized curve (orange) has smoothed out the divergency at $k=+k_0$ and completely removed the one at $k=-k_0$ due to the specific regularization path given in Eq.~\eqref{eq regularization}, also shown in  Fig.~\ref{figure 4}(b).

Figure ~\ref{figure 4}(d) presents the calculated  collective decay rate, which for the original $\omega_{\rm I}$ is constantly zero [cf. Fig.~\ref{figure 3}(b)], while the regularized curve shows a peak around $k=+k_0$.
Although the eigenvalues follow this peak fairly well, they also reveal a second peak at $k=-k_0$ similar to the real part, which is present neither in the original band structure $\omega_{\rm I}$, nor in the regularized one. 
The band structure in the complex plane is shown in Fig.~\ref{figure 4}(e), where $\omega_{\rm I}$ is a flat curve covering the real axis [cf. Fig.~\ref{figure 3}(c)], while the regularized curve forms a closed loop enclosing all finite system eigenvalues and providing all of them with a winding number $\nu(\varepsilon)=1$.
The non-trivial point gap topology of the regularized curve has been used in Ref.~\cite{poddubny2024} to justify the emergence of the non-Hermitian skin effect in this model.

The regularization leads to a fundamental change in the band structure, as shown in Fig.~\ref{figure 4}(e), such that the new band structure effectively describes a physical system different from the original one. 
This becomes even more evident when computing the couplings $t_n$ from Eq.~\eqref{eq inverse couplings} using the regularized band structure. 
The result is presented and compared to the original couplings in Fig.~\ref{figure 4}(f).
The red crosses show the absolute values of the original coupling strengths from Eq.~\eqref{eq couplings waveguide}, while the orange dots present the same quantities obtained after regularization. 
We can see that for negative $n$, the coupling strengths are visibly different as the values obtained after regularization decay with $\sim\exp(0.08n)$ (fitted black solid line), effectively describing a lossy waveguide.
(For positive $n$, the regularized couplings decay as well. However, since the original couplings are vanishingly small, also the regularized ones are approximately zero.)  
One can also see at $n=0$ that the single emitter decay rate, $\gamma_0 = -2\mathrm{Im}(t_0)$, has changed and is now smaller than $\gamma_0$. 
In Fig.~\ref{figure 4}(g) the values of this new decay rate $\gamma_{0,\mathrm{reg}}$ are plotted as a function of the chain length $N$, showing that only in the limit of large $N$ the original decay rate is recovered.

\begin{figure*}
    \centering
    \includegraphics[width=\linewidth]{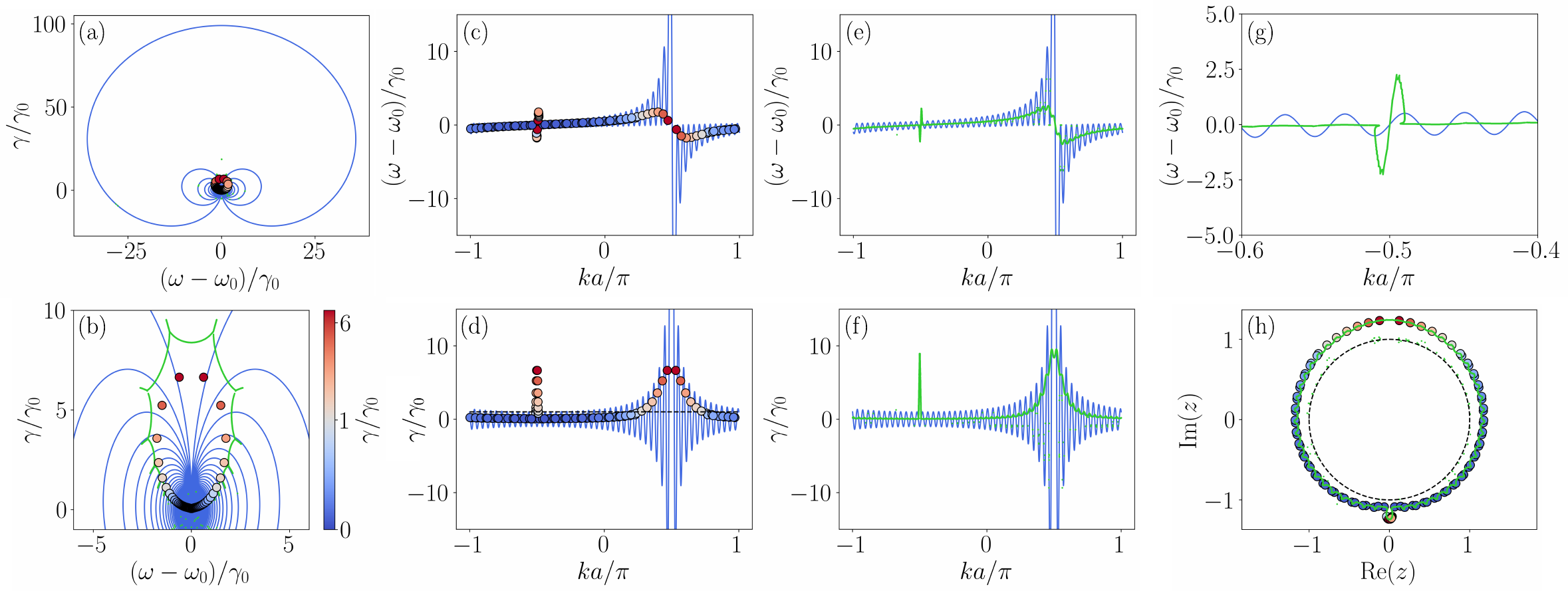}
    \caption{Band structure of a quantum emitter chain chirally coupled to a waveguide with  $\xi=2\times 10^{-5}$   calculated according to Method~II.
    (a) shows the band structure $\omega_{\rm II}$ as a curve (blue) in the complex plane. 
    In the close-up (b) we also present the calculated limiting set $\Lambda$ (green dots) and the finite size eigenvalues (colored dots where the color indicates the decay rate: red $\hat{=}$ superradiance, blue $\hat{=}$ subradiance).
    (c) and (d) show the frequency shift $\omega = \mathrm{Re}(\omega_{\rm II})$ and collective decay rates $\gamma = -2\mathrm{Im}(\omega_{\rm II})$ (blue curves), respectively, with the corresponding finite size eigenvalues [color code as in (b)].
    The dashed black line in (d) corresponds to $\gamma/\gamma_0=1$, marking the border between subradiance and superradiance.
    (e) and (f) show the same as (c) and (d), but including the limiting set $\Lambda$ (green) instead of the finite size eigenvalues. 
    (g) is a close-up of (e) around $k=-k_0$ showing the multivaluedness of $k \mapsto \Lambda(k)$.
    (h) Generalized Brillouin zone (green solid line) with the $z$ values corresponding to each eigenvalue [colored dots, same color code as in (b)].
    The black dashed line shows the conventional Brillouin zone $\mathrm{BZ} = \qty{z\in\mathbb{C} \, : \, |z|=1}$. Close-ups of this graph are presented in Appendix~\ref{sec appendix GBZ close ups}.
    Parameters used for all graphs are $N=50$, $\xi=2\times10^{-5}$, and $\varphi=\pi/2$.
    }
    \label{figure 5}
\end{figure*}

Since the regularization leads to couplings different from the ones in Eq.~\eqref{eq couplings waveguide}, we have to adjust our finite system Hamiltonian in Eq.~\eqref{eq quantum emitter Hamiltonian} using the regularized couplings for calculating the system eigenvalues, as we would otherwise produce inconsistencies. 
The regularized eigenvalues are shown as orange dots in Fig.~\ref{figure 4}(h) alongside with the original ones (red dots) and the regularized curve (orange curve).
One can see that the decaying couplings lead to a shrinking of the circle on which the eigenvalues are distributed.

The eigenstates of the regularized Hamiltonian show a similar localization to the one shown in Fig.~\ref{figure 4}(a), which can be satisfyingly explained by the non-trivial point-gap topology of the regularized band structure. 
However, since the regularized Hamiltonian describes a different physical system, the eigenstate localization of the original Hamiltonian still remains unexplained.

\subsection{Skin effect with Method II}
We now turn to Method~II and investigate whether it is suited to explain the emergence of skin states under chiral coupling. 
As mentioned at the beginning of this Section, this would require the band structure $\omega_{\rm II}$ to have a non-trivial point-gap topology, i.e., being a closed curve with a non-zero winding number $\nu(\varepsilon_0)$ for some $\varepsilon_0/\gamma_0 \in \mathbb{C}$.
In Fig.~\ref{figure 5}(a), we show the calculated band structure $\omega_{\rm II}(k)$ as a curve in the complex plane (blue curve).
Although this plot superficially looks similar to the one in Fig.~\ref{figure 3}(c) for the non-chiral case (which has no point gap), by numerically calculating the winding number one can verify that $\omega_{\rm II}$ indeed has a non-trivial point gap with a winding number $\nu=-1$ within the big outer loop and increasing winding numbers within the inner loops. 
According to the general theory \cite{Wang2024review, zhang2020winding, Okuma2020}, this justifies the eigenstate localization shown in Fig.~\ref{figure 4}(a).
Multi-loop band structures like the one in Fig.~\ref{figure 5}(a) have also been observed in Ref.~\cite{Wang2021science}, which uses a Hamiltonian distantly related to the one used here.

We continue our analysis by introducing two quantities that have proven themselves to be useful for understanding the non-Hermitian skin effect: the limiting set and the generalized Brillouin zone (GBZ).
The limiting set $\Lambda$ is defined as the eigenvalue spectrum $\sigma$ of the finite size Hamiltonian $H$ in Eq.~\eqref{eq quantum emitter Hamiltonian} when taking the limit of $N\to\infty$ \cite{toeplitz_book}:
\begin{equation}\label{eq limiting set}
    \Lambda := \lim_{N\to\infty} \sigma(H),
\end{equation}
where the limit is defined according to Method~II.
To compute $\Lambda$, we have used the algorithm presented in Ref.~\cite{boettcher2021}.
While in a Hermitian system the limiting set and the band structure curve coincide, in a non-Hermitian system with non-trivial point-gap topology they are generally different \cite{toeplitz_book}. 
Related to the limiting set is the GBZ. 
While the band structure $\omega_{\rm II}(\mathrm{BZ})$ is determined by the conventional Brillouin zone, $\mathrm{BZ}=\qty{z\in\mathbb{C} \, : \, |z|=1}$, the limiting set can be written as $\Lambda = \omega_{\rm II}(\mathrm{GBZ})$ \cite{Wang2024review}, showing that $\Lambda$ can be considered an alternative band structure notion for chiral systems.
The precise definition of the GBZ can be found in Appendix~\ref{appendix gbz and limiting set} and in more detail in Chapter~5 of Ref.~\cite{Wang2024review}.
We note that both the limiting set and the GBZ can only be defined in the framework of Method~II, since both notions require a band structure of the form given in Eq.~\eqref{eq omega short range}, of which Method~II is a special case.

In the close-up of the band structure $\omega_{\rm II}$ shown in Fig.~\ref{figure 5}(b), we also present the limiting set $\Lambda$ (green), which, in accordance with the general theory of the non-Hermitian skin effect, lies entirely in regions with winding number $\nu\neq0$ except for points where it crosses the band structure \cite{toeplitz_book, Wang2024review}.
The finite size eigenvalues obtained for a chain of $N=50$ emitters (colored dots) approximately follow the outline of the limiting set, as one expects from the definition of $\Lambda$ in Eq.~\eqref{eq limiting set}.

Figures~\ref{figure 5}(c) and (d) show the frequency shift $\omega(k)=\mathrm{Re}(\omega_{\rm II}(k))$ and collective decay rate $\gamma(k)=-2\mathrm{Im}(\omega_{\rm II}(k))$ as functions of $k$ together with the finite-size eigenvalues.
While the eigenvalues for the most part follow the band structure well with superradiant states localized around $k=+k_0$, there is a nearly vertical pile at $k=-k_0$ in both real and imaginary part, which is not present in the band structure. 
This is similar to our previous considerations with the regularized curve in Figs.~\ref{figure 4}(c) and (d).
While the band structure itself cannot explain this phenomenon, the limiting set and the generalized Brillouin zone can. 
By plotting the limiting set as a function of $k$ in Figs.~\ref{figure 5}(e) [$\mathrm{Re}(\Lambda)$] and (f) [$-2\mathrm{Im}(\Lambda)$], one can see that $\Lambda$ shows peaks in the same shape of the eigenvalue piles at $k=-k_0$, proving that this resonance peak only appears under open boundary conditions. 
The real part of this peak is particularly interesting as it shows that the map $k\mapsto\Lambda(k)$ is actually multivalued, cf. the small kinks in the close-up in Fig.~\ref{figure 5}(g). 
This means that in one-band models with asymmetric long-range couplings it is possible that one momentum $k$ is mapped to more than one eigenvalue.

The reason for this multivaluedness lies in the geometry of GBZ, which is shown in Fig.~\ref{figure 5}(h).
Since some important features are rather hard to identify in Fig.~\ref{figure 5}(h),  close-ups of the GBZ can be found in Appendix~\ref{sec appendix GBZ close ups}. 
While the conventional Brillouin zone BZ (black dashed line) encloses a star-shaped set, meaning that we can reach any point enclosed by BZ via a straight line starting from the origin, the GBZ (green) is non-star shaped due to the small ``bay'' at the bottom [see also Fig.~\ref{figure 9}(c) in Appendix~\ref{sec appendix GBZ close ups}], leading to the peculiarity that some points on the GBZ have the same complex phase and, therefore, the same $k$ value.
Such odd shapes of the GBZ are a topic currently attracting attention and even more obscure shapes, like non-connected GBZ's, have been studied in toy models \cite{Wang2024review, wang2026}. 
In Fig.~\ref{figure 5}(h), we also show the $z$ values corresponding to the eigenvalues of the finite size system, where the colors indicate the decay rate as in Fig.~\ref{figure 5}(b).
One can observe that the eigenvalues pairwise ``occupy'' the GBZ as long as the eigenstates are subradiant (i.e., $\gamma<\gamma_0$), see also the close-ups in Fig.~\ref{figure 9}(b) and (c) in Appendix~\ref{sec appendix GBZ close ups}. 
As soon as the eigenstates become superradiant ($\gamma>\gamma_0$), one half of their corresponding $z$ values are distributed along the top part of the GBZ, which corresponds to $k$ values around $+k_0$, while the other half sits densely packed in the small bay at the bottom, leading to the dense piles in Figs.~\ref{figure 5}(c) and (d) at around $k=-k_0$.

In summary, the Method~II band structure is fully capable of explaining the emergence of localized skin modes for the quantum emitter chain chirally coupled to a waveguide. 
Moreover, the unusual shape of the GBZ justifies the eigenvalue piles at $k=-k_0$, which could not be explained by the band structure itself (and remained entirely unexplained with Method~I).

\subsection{Skin effect with weak chirality}\label{sec skin effect with weak chirality}
\begin{figure}
    \centering
    \includegraphics[width=\linewidth]{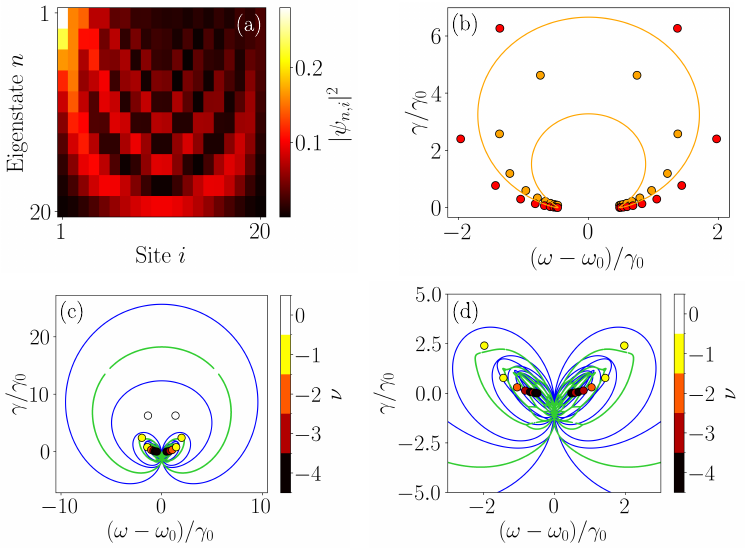}
    \caption{Non-Hermitian skin effect with chirality $\xi=0.5$.
    (a) Localization of eigenstates $\ket{\psi_n}$ of the finite size Hamiltonian in Eq.~\eqref{eq quantum emitter Hamiltonian} with $N=20$.
    Eigenstates are sorted by the negative imaginary part of their corresponding eigenvalues (i.e., $n=1$ corresponds to highest decay rate $\gamma$, $n=N$ to the lowest).
    (b) Regularized band structure from Method~I (orange curve) with eigenvalues of the original Hamiltonian (red dots) and the regularized one (orange dots).
    (c) Band structure from Method~II (blue curve) alongside the limiting set (green) and the finite size eigenvalues (colors mark the value of the winding number of each eigenvalue).
    (d) Close-up of (c).
    Parameters in all panels are $N=20$, $\xi=0.5$, and $\varphi=\pi/2$.}
    \label{figure 6}
\end{figure}
So far, we have studied the non-Hermitian skin effect in the waveguide system with a rather strong asymmetry of $\xi=2\times10^{-5}$ (implying $\gamma_{\rm L}\approx\gamma_0$ and $\gamma_{\rm R}\approx0$), a value which has been used before in Refs.~\cite{fedorovich2022, sheremet2023} since this leads to a very distinct localization. 
To provide a full picture, in the following we are going to compare the two considered Methods by analyzing the skin effect for a weaker chirality with directionality parameter $\xi=\frac{1}{2}$ (i.e., $\gamma_{\rm L} = 2\gamma_{\rm R}$).
In Fig.~\ref{figure 6}(a), we show the eigenstate localization for a finite size chain with $N=20$ emitters, where the eigenstates are sorted by decreasing collective decay rate.
(The shorter chain length has been chosen due to the complexity of the limiting set, see below.)
One can see that, compared to Fig.~\ref{figure 4}(a), only the first few eigenstates are visibly localized, while states with index $n>6$ are mostly delocalized and the most subradiant states appear as perfect Bloch modes.

Figure~\ref{figure 6}(b) shows the regularized Method~I band structure (cf.~Sec.~\ref{sec Skin effect with method I}).
Similar to before, the regularized couplings decay with distance $n$ approximately as $\sim \exp(-|n|/5)$, fundamentally changing our physical system. 
The eigenvalues of the correspondingly regularized Hamiltonian are shown as orange dots and the ones of the original Hamiltonian as red dots in Fig.~\ref{figure 6}(b), showing again that the additional decay leads to a shrinking of the ``circle'' on which the eigenvalues lie. 
Interestingly, although it has been shown in Ref.~\cite{poddubny2024} that the regularized curve encloses all (red) eigenvalues for $N=50$, our example proves that this is no longer the case for $N=20$ as here all red eigenvalues lie outside the orange curve. 
The regularized (orange) eigenvalues, however, for the most part are encircled by the band structure, except for the most subradiant ones. 
Again, the non-trivial point-gap topology of the regularized curved only justifies the localization of eigenstates of the regularized Hamiltonian, which corresponds to a different physical system. The localization of the regularized eigenstates looks similar to the one in Fig.~\ref{figure 6}(a).

The eigenstate localization is analyzed using Method~II in Figs.~\ref{figure 6}(c) and (d), where the blue curve shows the band structure and the green line is the limiting set (holes are  numerical artifacts).
The colored dots are the eigenvalues of the finite size Hamiltonian and the color indicates their individual winding numbers, which otherwise are hard to identify with the bare eye due to the complexity of the loop. 
The limiting set  has a rather complex structure, the beauty of which can be appreciated in Appendix~\ref{sec appendix limiting set}.

The calculated winding numbers presented in Fig.~\ref{figure 6}(c)  not only demonstrate the non-trivial point-gap topology; the two most subradiant states (white dots) reveal by their winding number of zero that the blue curve changes direction at some point such that these two points are encircled once clockwise and once counterclockwise. 
This leads to a total winding number of zero and, hence, these two states topologically lie outside of the blue loop, despite the fact that these are the ones that are the most localized in Fig.~\ref{figure 6}(a).

The strong localization of modes with zero winding and the delocalized states with non-zero winding numbers seemingly contradict the established non-Hermitian bulk-boundary correspondence. 
This concept  implies that under open boundary conditions (OBC) eigenvalues with non-zero winding number correspond to skin eigenstates \cite{zhang2020winding, Okuma2020}. 
However, this is only true in the limit of infinitely long chains, i.e., for the limiting set, not for the finite-size eigenvalues. 
This is the main reason why in this work the notion of limiting set is used (instead of the commonly used ``OBC spectrum''), as 
the term ``OBC spectrum'' does not directly distinguish between finite-size and infinite chains. 
While for the limiting set the one-to-one correspondence between winding number and skin states is well established in both physics and mathematics literature \cite{Okuma2020, zhang2020winding, toeplitz_book, Wang2024review}, for finite-size eigenvalues there is, to our knowledge, no such unambiguous correspondence. 
We believe that the simultaneous appearance of zero winding and strongly localized eigenstates originates from a combination of long-range couplings and finite system size, see also the complementary discussion in Appendix \ref{sec appendix zero winding}.

\section{Comparison}\label{sec comparison}
Having analyzed the two Methods on the basis of our quantum emitter models, it is now time to directly compare their strengths and weaknesses.  
Method~I has an obvious advantage over Method~II, which might be one reason why it has been used in a number of previous works, see Refs.~\cite{Garcia2017, fedorovich2022, poddubny2024, sheremet2023, Yu2022, Zeng2023, Calajo2022, Albrecht2019}:
The band structure obtained by Eq.~\eqref{eq omega 1} is independent of the size $N$ of the finite system. 
Therefore, given that $\omega_{\rm I}$ yields meaningful results, it can be applied for \textit{any} finite system size. 
In contrast to this, the band structure from Method~II in Eq.~\eqref{eq omega 2} has to be calculated individually for each system size.

That being said, Method~I has some serious disadvantages, which have become evident in our analysis. 
The most apparent issue is that the series in Eq.~\eqref{eq omega 1} is both physically and mathematically doubtful. 
Although an infinitely long chain is always a stretch of nature, all-to-all couplings in such a chain necessarily lead to divergencies in energy if the couplings decay too slowly. 
This happens already in the free-space chain with perpendicular polarization [cf. Eq.~\eqref{eq omega I freespace perpendicular}] and even more so in the waveguide model, where the couplings do not decay at all. 
The physical divergence of energy leads to the mathematical divergence of the series in Eq.~\eqref{eq omega 1}, which then can be found in the divergencies of the final results in Eq.~\eqref{eq omega I freespace perpendicular} and \eqref{eq omegaI waveguide}, respectively. 
These divergencies then inevitably lead to the inconsistency that the couplings cannot be recovered from the band structure using Eq.~\eqref{eq inverse couplings}, making it doubtful if $\omega_{\rm I}$ is actually capable of giving a proper description of the bulk spectrum.
Such issues do not occur in Method~II as the definition in Eq.~\eqref{eq omega 2} will always yield a smooth well-defined band structure.

The comparison of the band structures to the finite-size eigenvalues reveals more discrepancies. 
Although one cannot expect the eigenvalues to perfectly fit the band structure, they have to match up to a certain degree as otherwise the band structure would not properly describe the bulk energies. 
While for free-space quantum emitters, we find a fairly good agreement with the finite-size eigenvalues for both Methods, in the waveguide system there are severe discrepancies for Method~I.
The band structure in Eq.~\eqref{eq omegaI waveguide} has vanishing collective decay, which will never happen in a finite system due to conservation of the mean collective decay rate, cf. Fig.~\ref{figure 4}(e).
Therefore, $\omega_{\rm I}$ is unable to give a proper description of the bulk states. 
Contrary to this, $\omega_{\rm II}$ has a properly normalized collective decay rate and matches the behavior of the finite-size eigenvalues well, cf. Figs.~\ref{figure 5}(c) and (d).

We have also compared the two Methods based on their ability to explain the non-Hermitian skin effect that appears in the waveguide system with chiral couplings. 
We have seen that, due to the vanishing collective decay rate in $\omega_{\rm I}$, Method~I cannot readily explain the emergence of skin modes, which requires a non-trivial point gap. 
Although a regularization of the band structure can open such a gap, it also changes the physical system as the couplings that correspond to the regularized band structure differ from the original ones. 
Therefore, we conclude that Method~I is not able to deliver a satisfying explanation for the non-Hermitian skin effect. 
In contrast to this, the band structure obtained with Method~II has a proper point gap and also allows for the definition of the limiting set and the generalized Brillouin zone, which are central concepts in the theory of the non-Hermitian skin effect.

In conclusion, it becomes evident from our analysis that Method~II is much more convincing way for obtaining a meaningful band structure. 
There are no mathematical or physical pitfalls, it always yields a smooth and consistent result, and it is compatible with existing theory of the non-Hermitian skin effect. 
Still, Method~I has a right to exist. 
We have seen in the case of free-space emitters with parallel polarization that the two Methods yield almost perfectly agreeing results, which also fit well the finite-size eigenvalues. 
The main difference between this example and the perpendicular case, as well as the waveguide system, is that the couplings decay sufficiently fast ($\sim 1/n^2$) such that for a chain of $N=50$, the two formulas in Eqs.~\eqref{eq omega 1} and \eqref{eq omega 2} coincide [$t_{N-1} \sim 1/(N-1)^2 \approx 4\times 10^{-4}$].
We conclude that in systems that decay fast enough (faster than $1/n$), where the series in Eq.~\eqref{eq omega 1} converges, the two Methods are equivalent if $N$ is sufficiently large. 
In this case, Method~I would be slightly more attractive as it may allow for analytical results and is valid for any suitable system size.

\section{Summary}\label{sec summary}
While in tight-binding models with short-range couplings the band structure can be calculated from Eq.~\eqref{eq omega short range}, in the case of long-range couplings there are two distinct procedures, one of them using an infinite chain with all-to-all couplings (Method~I) and the other one extending the chain while keeping the finite range couplings (Method~II), which is also related to periodic boundary conditions. 
We have presented a detailed analysis of these two Methods by comparing them on the basis of quantum emitter models. The latter provide  several different forms of long-range couplings, which is why we believe that our results are also applicable to a broader class of systems.
We have found that in general Method~II is preferable over Method~I since it is mathematically reliable and physically consistent both in itself and with other concepts like the non-Hermitian skin effect. 
Only in cases in which the couplings decay fast enough and the finite system in sufficiently large, Method~I is slightly advantageous as its result is independent of the system size.

As a general remark, we want to point out that in systems with long-range couplings, one has to be careful with the notions of ``infinite chain'' and ``periodic boundary conditions'' as these generally are not the same. 
For instance, Ref.~\cite{poddubny2024} denotes its band structure as ``PBC'', although it has been obtained with Method~I.
As shown in Appendix~\ref{sec PBC}, periodic boundary conditions are equivalent to Method~II.

In our analysis of the non-Hermitian skin effect using Method~II, we have discovered a number of exciting peculiarities. 
Due to the non-star shaped structure of the generalized Brillouin zone under strong chirality, the limiting set becomes multivalued, leading to a pile up of finite-size eigenvalues at one of the resonance points. 
While odd shapes of the GBZ have been discussed in toy models \cite{wang2026}, our analysis shows that they can be found in realistic systems as well.
Moreover, in the case of weak chirality we have seen that some of the eigenvalues can even lie outside regions with non-zero winding number and still be strongly localized. 
This shows that there is no general correspondence between the winding number of finite-size eigenvalues and the localization of their corresponding eigenstates.
Such a correspondence has only been proven for the limiting set \cite{Okuma2020, zhang2020winding, toeplitz_book} and needs to be further studied for finite system sizes. 
We have provided a number of thoughts on this future subject in Appendix~\ref{sec appendix zero winding} and are confident that these observations will initiate further development in the field of non-Hermitian physics.

%-------- continue here! ........................

\begin{acknowledgments}
The authors gratefully
acknowledge financial support by the Deutsche Forschungsgemeinschaft (DFG, German Research Foundation) through the Würzburg-Dresden Cluster of Excellence ctd.qmat – Complexity, Topology and Dynamics in Quantum Matter (EXC 2147, project-id 390858490).
AP gratefully acknowledges the Heisenberg Program of the
DFG  (project-id 435041839).
\end{acknowledgments}

\appendix
\section{Periodic boundary conditions}\label{sec PBC}
In this Appendix, we prove that the Method~II band structure in Eq.~\eqref{eq omega 2} is consistent with the band structure obtained from periodic boundary conditions (PBC). 
We start from our initial Hamiltonian in Eq~\eqref{eq hamiltonian initial}:
\begin{equation}
    H = \sum_{i=1}^N \sum_{n=-i+1}^{N-i} t_n c_{i+n}^\dagger c_i,
\end{equation}
which under PBC ($t_{n\pm N}\equiv t_n$) becomes
\begin{equation}
    \Tilde{H} = \sum_{i=1}^N \sum_{n=-(N-1)}^{N-1} \Tilde{t}_n c_{i+n}^\dagger c_i,
\end{equation}
where $c_{i+N}^{(\dagger)} = c_i^{(\dagger)}$ and $\Tilde{t}_n$ are the components $\Tilde{t}_{ij} \equiv \Tilde{t}_{i-j} \equiv \Tilde{t}_n$ of a matrix $\Tilde{T}$ and read
\begin{equation}\label{eq tilde tn}
    \Tilde{t}_n = t_n +
    \begin{cases}
        t_{n-N} & (n\neq0), \\
        0 & (n=0).
    \end{cases}
\end{equation}
Explicitly, the matrix $\Tilde{T}$ generated by the $\Tilde{t}_n$ reads
\begin{equation}
    \Tilde{T} = 
    \mqty(\Tilde{t}_0 & \Tilde{t}_{N-1} & \Tilde{t}_{N-2} & \dots & \Tilde{t}_{1} \\
    \Tilde{t}_1 & \Tilde{t}_0 & \Tilde{t}_{N-1} & \dots & \Tilde{t}_{2} \\
    \Tilde{t}_2 & \Tilde{t}_1 & \Tilde{t}_0 & \dots & \Tilde{t}_3 \\
    \vdots & \vdots & \vdots & \ddots & \vdots \\
    \Tilde{t}_{N-1} & \Tilde{t}_{N-2} & \Tilde{t}_{N-3} & \dots & \Tilde{t}_0
    ).
\end{equation}
The spectrum of such a circulant matrix is given by \cite{toeplitz_book}
\begin{equation}
    \sigma(\Tilde{T}) = \qty{\Tilde{\omega}\qty(\eu^{\imi k_0}), \ \Tilde{\omega}\qty(\eu^{\imi k_1}),\ ..., \ \Tilde{\omega}\qty(\eu^{\imi k_{N-1}})}, 
\end{equation}
where $k_n = \frac{2\pi n}{N}$ ($n=0,...,N-1$) are the discrete momenta and
\begin{equation}
    \Tilde{\omega}(z) = \sum_{n=0}^{N-1} \Tilde{t}_n z^n
\end{equation}
can be interpreted as the band structure under PBC.
Using the definition of the couplings $\Tilde{t}_n$ in Eq.~\eqref{eq tilde tn}, 
we find that 
\begin{align}
    \Tilde{\omega}(z) & = t_0 + \sum_{n=1}^{N-1} t_n z^n + \sum_{n=1}^{N-1} t_{n-N} z^n \nonumber \\
    & =  t_0 + \sum_{n=1}^{N-1} + \sum_{n=-(N-1)}^{-1} t_{-n-N} z^n \nonumber \\
    & = \sum_{n=-(N-1)}^{N-1} t_n z^n = \omega_{\rm II}(z),
\end{align}
where in the last step we used that under PBC we have $t_{-n-N} = t_{-n}$.
This proves that the Method~II band structure is the continuum version of the PBC band structure.

\section{Definition of the GBZ and limiting set}\label{appendix gbz and limiting set}
In this Appendix, we provide the definition of the generalized Brillouin zone GBZ and the limiting set. 
We start from the band structure in Eq.~\eqref{eq omega short range}
\begin{equation}
    \omega(z) = \sum_{n=-p}^{q} t_n z^n,
\end{equation}
with $p,q\in\mathbb{N}$, which also covers Method~II (with $p=q=N-1$).
For any $\varepsilon\in\mathbb{C}$, the polynomial 
\begin{equation}
    P_\varepsilon(z) = z^p(\omega(z) - \varepsilon)
\end{equation}
has $p+q$ zeros (fundamental theorem of algebra), which we call $z_1(\varepsilon),...,z_{p+q}(\varepsilon)$, sorted by increasing magnitudes:
\begin{equation}
    |z_1(\varepsilon)| \leq ... \leq |z_{p+q}(\varepsilon)|.
\end{equation}
The GBZ is then defined as \cite{Wang2024review} 
\begin{equation}
    \mathrm{GBZ} = \qty{z_p, z_{p+1}\in\mathbb{C} \, : \, |z_p(\varepsilon)|=|z_{p+1}(\varepsilon)|}.
\end{equation}
One can then show that the limiting set is given by \cite{toeplitz_book, Wang2024review, Schmidt1960}
\begin{equation}
    \Lambda = \omega(\mathrm{GBZ}) = \qty{\varepsilon\in\mathbb{C} \, : \, |z_p(\varepsilon)| = |z_{p+1}(\varepsilon)|}.
\end{equation}
From this expression, one can derive the algorithm that we have used for computing the limiting set, see Ref.~\cite{boettcher2021}.

\section{Mapping eigenvalues to $k$ values}\label{appendix eigenvalue mapping}
This Appendix explains how we mapped the eigenvalues of finite size Hamiltonians to $k$ values on the (generalized) Brillouin zone, as it was done in Figs.~\ref{figure 4}(c) and (d) for Method~I and in Figs.~\ref{figure 5}(c) and (d) for Method~II. \\

\textit{Method~I.}
For Method~I, we start from the band structure $\omega_{\rm I}$ given in Eq.~\eqref{eq omegaI waveguide}, which for the case $\varphi=\pi/2$ takes the form 
\begin{equation}
    \omega_{\rm I}(z) = \omega_0 + \gamma_0 \frac{1 + \frac{\chi}{2\imi} (z - z^{-1})}{z + z^{-1}}, 
\end{equation}
with $\chi = \frac{1-\xi}{1+\xi}$.
Solving the Equation $\omega_{\rm I}(z) = \varepsilon$ for some frequency $\varepsilon$ yields the solutions \cite{poddubny2024}
\begin{equation}
    z_\pm(\varepsilon) = \frac{1 \pm \sqrt{1 - \chi^2 - 4\Omega^2}}{\imi\chi + 2\Omega}, 
\end{equation}
with $\Omega = (\varepsilon - \omega_0)/\gamma_0$.
To each eigenvalue $\varepsilon$ of the finite-size Hamiltonian, we can now assign two $k$ values given by $k_\pm(\varepsilon) = \mathrm{arg}(z_\pm(\varepsilon))$. \\

\textit{Method~II.}
For Method~II one could use a similar strategy as above by solving the Equation $\omega_{\rm I}(z) = \varepsilon$ for each eigenvalue $\varepsilon$, yielding solutions $z_1, ..., z_{2N-2}$, which we sort by increasing magnitude. 
The $k$ values would then be given by $k_1(\varepsilon)= \mathrm{arg}(z_{N-1}(\varepsilon))$ and $k_2 = \mathrm{arg}(z_N(\varepsilon))$.

However, we have found that a more robust way is to first map each eigenvalue to the element $\lambda_\varepsilon\in\Lambda$ of the limiting set with the smallest distance to $\varepsilon$.
We then use the two $z_{1,2}\in\mathrm{GBZ}$ with $\omega_{\rm II}(z_{1,2}) = \lambda_\varepsilon$ and set $k_{1,2} = \mathrm{arg}(z_{1,2})$.
This way, we make sure that each $k$ value is directly extracted from the GBZ.

\section{Details of the GBZ for strong chirality}\label{sec appendix GBZ close ups}
\begin{figure}
    \centering
    \includegraphics[width=\linewidth]{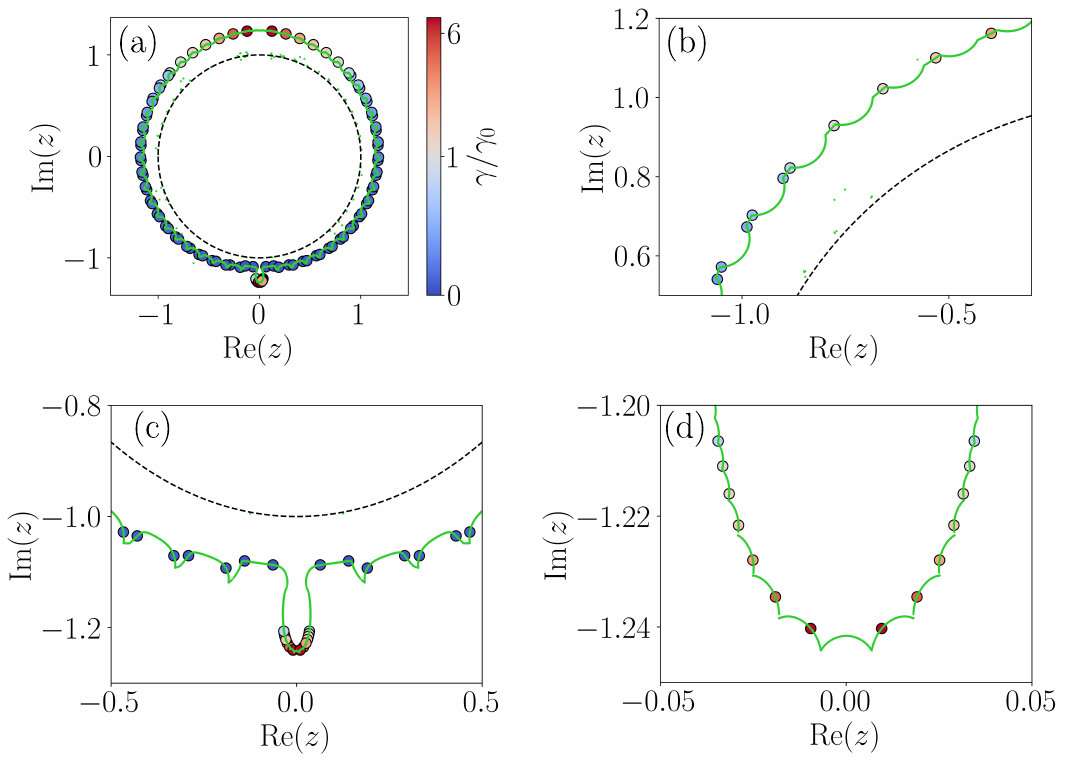}
    \caption{Generalized Brillouin zone (green) with various close-ups.
    The black dashed line is the conventional Brillouin zone and the dots are the $z$ values corresponding to the finite size eigenvalues.
    The color code indicates the collective decay rate of each eigenvalue.
    Parameters for all graphs are $N=50$, $\xi=2\times10^{-5}$, and $\varphi=\pi/2$.}
    \label{figure 9}
\end{figure}
Figure~\ref{figure 9} presents the generalized Brillouin zone GBZ that we have discussed in Fig.~\ref{figure 5}(h) in the main text with various close-ups. 
The dots are the $z$ values corresponding to the eigenvalues of the finite size Hamiltonian (in the following simply called $z$ values) and the color code indicates their collective decay rates. 
In Fig.~\ref{figure 9}(a), we see that the GBZ is mostly an enlarged version of the conventional BZ. 
In the close-up in Fig.~\ref{figure 9}(b), however, one observes that the green circle actually has sun-like prongs, which are occupied by the $z$ values. 
As long as the eigenstates are subradiant ($\gamma<\gamma_0$), the prongs are double occupied, whereas for superradiant states ($\gamma>\gamma_0$), each prong is occupied by a single $z$ value. 

An unusual feature is the small ``bay'' at the bottom of the GBZ, which can be seen in Fig.~\ref{figure 9}(c).
The specific shape of this bay, which looks similar to an upside down Greek $\Omega$ leads to the multivaluedness of the limiting set $\Lambda$, which is shown in Fig.~\ref{figure 5}(g) in the main text. 
As seen in Figs.~\ref{figure 9}(c) and (d), the bay is densely occupied by superradiant $z$ values (these are the ones that are missing at the top of the GBZ), leading to the dense eigenvalue piles in Figs.~\ref{figure 5}(c) and (d).

\section{Limiting set for $\xi=0.5$}\label{sec appendix limiting set}
\begin{figure}
    \centering
    \includegraphics[width=\linewidth]{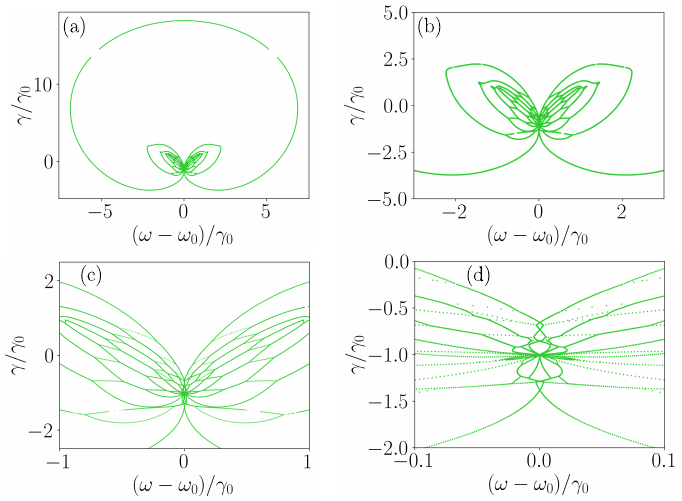}
    \caption{Limiting set $\Lambda$ with various zooms. 
    Parameters are $N=20$, $\xi=0.5$, and $\varphi=\pi/2$.}
    \label{figure 7}
\end{figure}
In Fig.~\ref{figure 7}, we present the limiting set for the $\xi=0.5$ case in a chain of $N=20$ emitters, which has been discussed in the main text in Figs.~\ref{figure 6}(c) and (d).
By increasing the zoom onto the origin in Figs.~\ref{figure 7}(b) and (c), one can observe a rich structure resembling the leaves of a flower or the wings of a butterfly.
Although it is an established fact that the limiting set can be multiply connected (see, e.g., Theorem~11.20 in Ref.~\cite{toeplitz_book} or the recent work on disconnected generalized Brillouin zones in Ref.~\cite{wang2026}), the high number of regions separated by $\Lambda$ is surprising. 
For instance, Theorem~11.20 in Ref.~\cite{toeplitz_book} proves that for specific types of band structures, one can expect the complex plane to be separated into around $N/2$ regions, where $N$ is the chain length. 
The band structure considered here does not fulfill the prerequisites of Theorem~11.20 in Ref.~\cite{toeplitz_book}, yet we find it astonishing that the limiting set calculated here divides the plane into far more than $N$ regions.

\section{Comments on skin modes with zero winding}\label{sec appendix zero winding}
\begin{figure*}
    \centering
    \includegraphics[width=\linewidth]{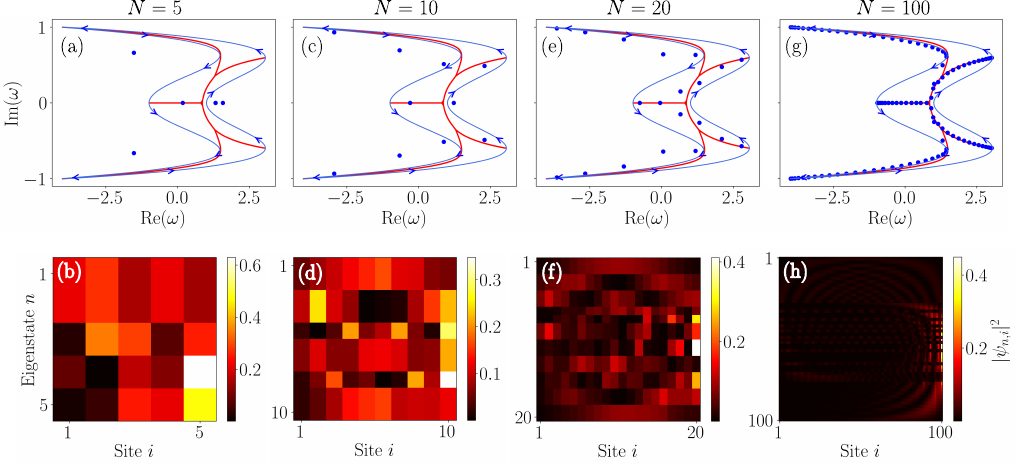}
    \caption{Band structure and eigenstates for the couplings given in Eq.~\eqref{eq appendix couplings} for various system sizes $N$.
    The top row shows the band structure loop (blue curve), where the arrows indicate the winding direction. 
    Blue dots are the finite size eigenvalues and the red curve is the limiting set.
    The bottom row shows the eigenstate localization $|\psi_{n,i}|^2$ of eigenstates corresponding to the $n^\mathrm{th}$ eigenvalue sorted by increasing real part.}
    \label{figure 10}
\end{figure*}
This Appendix complements the discussion on skin modes with zero winding number from Sec.~\ref{sec skin effect with weak chirality}. 
We have seen in Fig.~\ref{figure 6} that there can be states with zero winding number that are yet strongly localized, as well as states with finite winding that are delocalized. 
To make sure that this is not just a peculiarity of our waveguide model when using Method II, or a numerical error, we show that one can observe such features also in simple toy models with short-range couplings.

Consider the Hamiltonian from Eq.~\eqref{eq hamiltonian initial},
\begin{equation}
    H = \sum_{i=1}^N \sum_{n = -i+1}^{N-i} t_n c_{i+n}^\dagger c_i,
\end{equation}
and the couplings 
\begin{equation}\label{eq appendix couplings}
    t_{-4} = t_4 = -1, \ \ \ t_{-2} = t_1 = t_2 = 1,
\end{equation}
and $t_n=0$ else.
Using Eq.~\eqref{eq omega short range}, we obtain the band structure
\begin{equation}
    \omega(z) = -z^{-4} + z^{-2} + z + z^2 - z^4,
\end{equation}
which is shown as the blue curve in the top row of Fig.~\ref{figure 10}, where the arrows indicate the direction of the curve winding.
We can see that, as expected from the general theory \cite{Okuma2020, zhang2020winding, Wang2024review, toeplitz_book}, the limiting set $\Lambda$ (red) is fully enclosed by the band structure loop, such that each point of $\Lambda$ has a non-zero winding number $\nu = 1$, possibly except for single points that lie on top of $\omega$.

However, the finite-size eigenvalues (blue dots) are generally not all enclosed by $\omega$.
For $N=5$, which is a very short chain compared to the coupling length, all eigenvalues except for one lie outside the loop, see Fig.~\ref{figure 10}(a).
The eigenstates (ordered from top to bottom by increasing real part of the eigenvalues) shown in Fig.~\ref{figure 10}(b) reveal that the one eigenstate with non-zero winding (as well as the first two eigenstates) is delocalized, while eigenstates 4 and 5, which have zero winding, are clearly localized to the right.
A similar behavior can be observed for $N=10$, which is shown in Figs.~\ref{figure 10}(c) and (d).
Some of the eigenstates with non-zero winding (such as the last two states) are rather delocalized, while eigenstate $n=7$, which has zero winding, is strongly localized.

When further increasing the chain length to $N=20$ in Figs.~\ref{figure 10}(e) and (f), an increasingly large portion of the eigenvalues lie inside the loop, starting to follow the shape of the limiting set. 
The one skin state with winding number zero that we have seen for $N=10$ is not anymore present for $N=20$. 
Instead, the two most localized states correspond to the two eigenvalues that lie close to the center of the enclosed region. 
For long chains, the eigenvalues follow the limiting set well, as one can see in Fig.~\ref{figure 10}(g). 
However, even there several points lie outside the loop, namely in regions where the limiting set closely follows the shape of the band structure. 
Figure~\ref{figure 10}(h) shows that eigenvalues located in regions where the loop is wide are strongly localized, whereas eigenvalues in the thinner areas are delocalized.

From these observations, we draw the following conclusions: 
\begin{enumerate}
    \item Finite-size eigenvalues do not need to be encircled by the band structure. 
    Only points of the limiting set must lie within the band structure loop.
    \item The winding number of finite-size eigenvalues has no immediate implication on the localization of the corresponding eigenstates. 
    States with zero winding can be strongly localized, while states with finite winding can be fully delocalized. 
    \item For increasing lattice size $N$, the finite-size eigenvalues move towards the limiting set. 
    This follows directly from the definition of $\Lambda = \lim_{N\to\infty} \sigma(H)$.
    Therefore, zero-winding eigenstates are a consequence of finite structure size. 
    The further the coupling range is, the longer the chain needs to be for the eigenvalues to approach the limiting set. 
    \item The localization strength of eigenstates seems to depend on the wideness of the band structure loop in the region where the corresponding eigenvalues are located. 
    This can be seen in Figs.~\ref{figure 10}(g) and (h), where eigenvalues with negative real part lie in very narrow regions of the loop such that their corresponding eigenstates are delocalized. 
    In the same way, eigenvalues that lie further away from the band structure seem to be more localized. 
    \item The localization strength of points with winding number zero seems to depend on how they converge to the limiting set for $N\to\infty$. 
    The two points with zero winding and strong localization in Figs~\ref{figure 10}(a) and (b) move towards the central part of the limiting set, where the loop is rather wide. 
    Similarly, the two points in Fig.~\ref{figure 10}(a) with zero winding whose eigenstates in Fig.~\ref{figure 10}(b) are delocalized move towards the thin regions of the band structure. 
\end{enumerate}
However, these conclusions are of empirical nature and require more rigorous treatment. 
One difficulty here is that for the zero-winding skin states one needs to study the convergence behavior of the finite-size eigenvalues. 
However, this convergence is defined in terms of convergence of \textit{sets}, meaning that the set $\sigma(H)$ of finite-size eigenvalues converges to the set $\Lambda$. 
Therefore, it is not possible to directly study the convergence of a single eigenvalue. 
Rigorously proving the above statements goes far beyond the scope of this work and we leave these points as observations for now.

\newpage
\bibliography{bib.bib}% Produces the bibliography via BibTeX.

@PREAMBLE{
 "\providecommand{\noopsort}[1]{}" 
 # "\providecommand{\singleletter}[1]{#1}%" 
}

@book{kittel2018,
  title = {Introduction to {{Solid State Physics}}},
  author = {Kittel, Charles},
  edition = {8th},
  year = {2018},
  publisher = {John Wiley \& Sons},
  isbn = {978-1118838396}
}

@article{Kunst2018,
  title = {{Biorthogonal Bulk-Boundary Correspondence in Non-Hermitian Systems}},
  author = {Kunst, Flore K. and Edvardsson, Elisabet and Budich, Jan Carl and Bergholtz, Emil J.},
  journal = {Phys. Rev. Lett.},
  volume = {121},
  issue = {2},
  pages = {026808},
  numpages = {6},
  year = {2018},
  month = {Jul},
  publisher = {American Physical Society},
  doi = {10.1103/PhysRevLett.121.026808},
  url = {https://link.aps.org/doi/10.1103/PhysRevLett.121.026808}
}

@article{Wang2024review,
author = {Heming Wang and Janet Zhong and Shanhui Fan},
journal = {Adv. Opt. Photon.},
pages = {659},
publisher = {Optica Publishing Group},
title = {{Non-Hermitian photonic band winding and skin effects: a tutorial}},
volume = {16},
year = {2024},
doi = {10.1364/AOP.529289}
}

@article{chiu2016,
  title = {{Classification of topological quantum matter with symmetries}},
  author = {Chiu, Ching-Kai and Teo, Jeffrey C. Y. and Schnyder, Andreas P. and Ryu, Shinsei},
  journal = {Rev. Mod. Phys.},
  volume = {88},
  issue = {3},
  pages = {035005},
  numpages = {63},
  year = {2016},
  month = {Aug},
  publisher = {American Physical Society},
  doi = {10.1103/RevModPhys.88.035005},
  url = {https://link.aps.org/doi/10.1103/RevModPhys.88.035005}
}

@misc{moegerle2026,
      title={{Accurate Modeling of Rydberg Atoms and Their Interactions: Theory and Implementation in PairInteraction}}, 
      author={Johannes Mögerle and Frederic Hummel and Alicia Keil and Tangi Legrand and Eduard J. Braun and Henri Menke and Jonathan King and Beatriz Olmos and Sebastian Hofferberth and Hans Peter Büchler and Sebastian Weber},
      year={2026},
      eprint={2605.14993},
      archivePrefix={arXiv},
      primaryClass={physics.atom-ph},
      url={https://arxiv.org/abs/2605.14993}, 
}

@article{Yu2022,
  title = {{Giant microwave sensitivity of a magnetic array by long-range chiral interaction driven skin effect}},
  author = {Yu, Tao and Zeng, Bowen},
  journal = {Phys. Rev. B},
  volume = {105},
  issue = {18},
  pages = {L180401},
  numpages = {6},
  year = {2022},
  month = {May},
  publisher = {American Physical Society},
  doi = {10.1103/PhysRevB.105.L180401},
  url = {https://link.aps.org/doi/10.1103/PhysRevB.105.L180401}
}

@article{Zeng2023,
  title = {{Radiation-free and non-Hermitian topology inertial defect states of on-chip magnons}},
  author = {Zeng, Bowen and Yu, Tao},
  journal = {Phys. Rev. Res.},
  volume = {5},
  issue = {1},
  pages = {013003},
  numpages = {12},
  year = {2023},
  month = {Jan},
  publisher = {American Physical Society},
  doi = {10.1103/PhysRevResearch.5.013003},
  url = {https://link.aps.org/doi/10.1103/PhysRevResearch.5.013003}
}

@article{Pientka2013,
  title = {{Topological superconducting phase in helical Shiba chains}},
  author = {Pientka, Falko and Glazman, Leonid I. and von Oppen, Felix},
  journal = {Phys. Rev. B},
  volume = {88},
  issue = {15},
  pages = {155420},
  numpages = {13},
  year = {2013},
  month = {Oct},
  publisher = {American Physical Society},
  doi = {10.1103/PhysRevB.88.155420},
  url = {https://link.aps.org/doi/10.1103/PhysRevB.88.155420}
}

@article{Ren2026,
author = {Ren, Juntong and Lü, Haifeng},
title = {{Long-Range Interactions in Topological Superconducting Systems: A Mini Review}},
journal = {Advanced Physics Research},
year = {2026},
pages = {e00240},
doi = {https://doi.org/10.1002/apxr.202500240},
}

@book{soukoulis_book,
title = {{Photonic Band Gap Materials}},
author = {Soukoulis (Ed.), C. M.},
publisher = {Springer Dordrecht},
year = {1996},
doi = {https://doi.org/10.1007/978-94-009-1665-4},
}

@book{toeplitz_book,
title = {{Spectral Properties of Banded Toeplitz Matrices}},
author = {Albrecht Böttcher and Sergei M. Grudsky},
publisher = {SIAM Philadelphia},
year = {2005},
doi = {https://doi.org/10.1137/1.9780898717853},
}

@article{Okuma2020,
  title = {{Topological Origin of Non-Hermitian Skin Effects}},
  author = {Okuma, Nobuyuki and Kawabata, Kohei and Shiozaki, Ken and Sato, Masatoshi},
  journal = {Phys. Rev. Lett.},
  volume = {124},
  issue = {8},
  pages = {086801},
  numpages = {7},
  year = {2020},
  month = {Feb},
  publisher = {American Physical Society},
  doi = {10.1103/PhysRevLett.124.086801},
  url = {https://link.aps.org/doi/10.1103/PhysRevLett.124.086801}
}

@article{zhang2020winding,
  title = {{Correspondence between Winding Numbers and Skin Modes in Non-Hermitian Systems}},
  author = {Zhang, Kai and Yang, Zhesen and Fang, Chen},
  journal = {Phys. Rev. Lett.},
  volume = {125},
  issue = {12},
  pages = {126402},
  numpages = {6},
  year = {2020},
  month = {Sep},
  publisher = {American Physical Society},
  doi = {10.1103/PhysRevLett.125.126402},
  url = {https://link.aps.org/doi/10.1103/PhysRevLett.125.126402}
}

@article{wang2026,
  title = {{Topology of the Generalized Brillouin Zone of One-Dimensional Models}},
  author = {Wang, Heming and Zhong, Janet and Fan, Shanhui},
  journal = {Phys. Rev. Lett.},
  volume = {136},
  issue = {20},
  pages = {206601},
  numpages = {8},
  year = {2026},
  month = {May},
  publisher = {American Physical Society},
  doi = {10.1103/7cnd-r7q4},
  url = {https://link.aps.org/doi/10.1103/7cnd-r7q4}
}

@article{Lee2018,
    author = {Lee, C. H. and Imhof, S. and Berger, C. and Bayer, F. and Brehm, J. and Molenkamp, L. W. and Kiessling, T. and Thomale, R.},
    title = {{Topolectrical Circuits}},
    journal = {Commun Phys},
    volume = {1},
    pages = {39},
    year = {2018},
    doi = {https://doi.org/10.1038/s42005-018-0035-2}
}

@article{defenu2023,
  title = {{Long-range interacting quantum systems}},
  author = {Defenu, Nicol\`o and Donner, Tobias and Macr\`{\i}, Tommaso and Pagano, Guido and Ruffo, Stefano and Trombettoni, Andrea},
  journal = {Rev. Mod. Phys.},
  volume = {95},
  issue = {3},
  pages = {035002},
  numpages = {70},
  year = {2023},
  month = {Aug},
  publisher = {American Physical Society},
  doi = {10.1103/RevModPhys.95.035002},
  url = {https://link.aps.org/doi/10.1103/RevModPhys.95.035002}
}

@article{Sturm2025,
  title = {{Polarization-dependent topology in quantum emitter chains}},
  author = {Sturm, Jonathan and P\'alffy, Adriana},
  journal = {Phys. Rev. Res.},
  volume = {7},
  issue = {3},
  pages = {L032069},
  numpages = {7},
  year = {2025},
  month = {Sep},
  publisher = {American Physical Society},
  doi = {10.1103/3t6s-fnsq},
  url = {https://link.aps.org/doi/10.1103/3t6s-fnsq}
}

@article{poddubny2024,
  title = {{Mesoscopic non-Hermitian skin effect}},
  author = {Poddubny, Alexander and Zhong, Janet and Fan, Shanhui},
  journal = {Phys. Rev. A},
  volume = {109},
  issue = {6},
  pages = {L061501},
  numpages = {6},
  year = {2024},
  month = {Jun},
  publisher = {American Physical Society},
  doi = {10.1103/PhysRevA.109.L061501},
  url = {https://link.aps.org/doi/10.1103/PhysRevA.109.L061501}
}

@article{sheremet2023,
  title = {{Waveguide quantum electrodynamics: Collective radiance and photon-photon correlations}},
  author = {Sheremet, Alexandra S. and Petrov, Mihail I. and Iorsh, Ivan V. and Poshakinskiy, Alexander V. and Poddubny, Alexander N.},
  journal = {Rev. Mod. Phys.},
  volume = {95},
  issue = {1},
  pages = {015002},
  numpages = {59},
  year = {2023},
  month = {Mar},
  publisher = {American Physical Society},
  doi = {10.1103/RevModPhys.95.015002},
  url = {https://link.aps.org/doi/10.1103/RevModPhys.95.015002}
}

@article{fedorovich2022,
  title = {{Chirality-driven delocalization in disordered waveguide-coupled quantum arrays}},
  author = {Fedorovich, Gleb and Kornovan, Danil and Poddubny, Alexander and Petrov, Mihail},
  journal = {Phys. Rev. A},
  volume = {106},
  issue = {4},
  pages = {043723},
  numpages = {11},
  year = {2022},
  month = {Oct},
  publisher = {American Physical Society},
  doi = {10.1103/PhysRevA.106.043723},
  url = {https://link.aps.org/doi/10.1103/PhysRevA.106.043723}
}

@article{Albrecht2019,
    author = {Albrecht, A. and Henriet, L. and Asenjo-Garcia, A. and Dieterle, P. B. and Painter, O. and Chang, D. E.},
    title = {{Subradiant states of quantum bits coupled to a one-dimensional waveguide}},
    journal = {New J. Phys.},
    volume = {21},
    pages = {025003},
    year = {2019},
    doi = {10.1088/1367-2630/ab0134}
}

@article{Calajo2022,
  title = {{Emergence of solitons from many-body photon bound states in quantum nonlinear media}},
  author = {Calaj\'o, G. and Chang, D. E.},
  journal = {Phys. Rev. Res.},
  volume = {4},
  issue = {2},
  pages = {023026},
  numpages = {19},
  year = {2022},
  month = {Apr},
  publisher = {American Physical Society},
  doi = {10.1103/PhysRevResearch.4.023026},
  url = {https://link.aps.org/doi/10.1103/PhysRevResearch.4.023026}
}

@article{Lodahl2017,
    author = {Lodahl, P. and Mahmoodian, S. and Stobbe, S. and Rauschenbeutel, A. and Schneeweiss, P. and Volz, J. and Pichler, H. and Zoller, P.},
    title = {{Chiral quantum optics}},
    journal = {Nature},
    volume = {541},
    pages = {473},
    year = {2017},
    doi = {https://doi.org/10.1038/nature21037}
}

@article{Garcia2017,
  title = {{Exponential Improvement in Photon Storage Fidelities Using Subradiance and ``Selective Radiance'' in Atomic Arrays}},
  author = {Asenjo-Garcia, A. and Moreno-Cardoner, M. and Albrecht, A. and Kimble, H. J. and Chang, D. E.},
  journal = {Phys. Rev. X},
  volume = {7},
  issue = {3},
  pages = {031024},
  numpages = {36},
  year = {2017},
  month = {Aug},
  publisher = {American Physical Society},
  doi = {10.1103/PhysRevX.7.031024},
  url = {https://link.aps.org/doi/10.1103/PhysRevX.7.031024}
}

@article{Garcia2017_A,
  title = {{Atom-light interactions in quasi-one-dimensional nanostructures: A Green's-function perspective}},
  author = {Asenjo-Garcia, A. and Hood, J. D. and Chang, D. E. and Kimble, H. J.},
  journal = {Phys. Rev. A},
  volume = {95},
  issue = {3},
  pages = {033818},
  numpages = {16},
  year = {2017},
  month = {Mar},
  publisher = {American Physical Society},
  doi = {10.1103/PhysRevA.95.033818},
  url = {https://link.aps.org/doi/10.1103/PhysRevA.95.033818}
}

@article{Reitz2022,
  title = {{Cooperative Quantum Phenomena in Light-Matter Platforms}},
  author = {Reitz, Michael and Sommer, Christian and Genes, Claudiu},
  journal = {PRX Quantum},
  volume = {3},
  issue = {1},
  pages = {010201},
  numpages = {33},
  year = {2022},
  month = {Jan},
  publisher = {American Physical Society},
  doi = {10.1103/PRXQuantum.3.010201},
  url = {https://link.aps.org/doi/10.1103/PRXQuantum.3.010201}
}

@article{boettcher2021,
    author = {Böttcher, A. and Gasca, J. and Grudsky, S. M. and Kozak, A. V.},
    title = {Eigenvalue Clusters of Large Tetradiagonal Toeplitz Matrices},
    journal = {Integr. Equ. Oper. Theory},
    volume = {93},
    pages = {8},
    year = {2021},
    doi = {https://doi.org/10.1007/s00020-020-02619-z}
}

@article{Schmidt1960,
    author = {Schmidt, P. and Spitzer, F.},
    title = {{The Toeplitz matrices of an arbitrary Laurent polynomial}},
    journal = {Math. Scand.},
    volume = {8},
    pages = {15},
    year = {1960},
    doi = {https://doi.org/10.7146/math.scand.a-10588 }
}

@article{Wang2021science,
author = {Kai Wang  and Avik Dutt  and Ki Youl Yang  and Casey C. Wojcik  and Jelena Vučković  and Shanhui Fan },
title = {{Generating arbitrary topological windings of a non-Hermitian band}},
journal = {Science},
volume = {371},
number = {6535},
pages = {1240},
year = {2021},
doi = {10.1126/science.abf6568}
}

@article{Hewitt1979,
author = {Edwin Hewitt and Robert E.~Hewitt},
title = {{The Gibbs-Wilbraham phenomenon: An episode in fourier analysis}},
journal = {Arch. Hist. Exact Sci.},
volume = {21},
pages = {129},
year = {1979},
doi = {https://doi.org/10.1007/BF00330404}
}

@article{Leseleuc2019,
author = {Sylvain de Léséleuc  and Vincent Lienhard  and Pascal Scholl  and Daniel Barredo  and Sebastian Weber  and Nicolai Lang  and Hans Peter Büchler  and Thierry Lahaye  and Antoine Browaeys },
title = {{Observation of a symmetry-protected topological phase of interacting bosons with Rydberg atoms}},
journal = {Science},
volume = {365},
number = {6455},
pages = {775-780},
year = {2019},
doi = {10.1126/science.aav9105},
}

\end{document}